\makeatletter
\def\cl@chapter{}
\makeatother
\documentclass[epj]{svjour}

\usepackage{xcolor}

\usepackage{url}
\usepackage[utf8]{inputenc}
\usepackage{graphicx}
\usepackage{braket}
\usepackage{xintexpr}
\usepackage{siunitx}
\usepackage{amsmath,amssymb,amsfonts}%
\usepackage{xspace}
\usepackage{hyperref}
\usepackage{cleveref}

\usepackage{multirow}%
\usepackage{mathrsfs}%
\usepackage[title]{appendix}%
\usepackage{xcolor}%
\usepackage{textcomp}%
\usepackage{manyfoot}%
\usepackage{booktabs}%
\usepackage{algorithm}%
\usepackage{algorithmicx}%
\usepackage{algpseudocode}%
\usepackage{listings}%

\DeclareMathOperator*{\argmax}{arg\,max}

\abstract{Axion-like particles (ALPs) that decay into photon pairs pose a challenge for experiments that rely on the construction of a decay vertex in order to search for long-lived particles. 
This is particularly true for beam-dump experiments, where the distance between the unknown decay position and the calorimeter can be very large.
In this work we use machine learning to explore the possibility to reconstruct the ALP properties, in particular its mass and lifetime, from such inaccurate observations. 
We use a simulation-based inference approach based on conditional invertible neural networks to reconstruct the posterior probability of the ALP parameters for a given set of events. 
We find that for realistic angular and energy resolution, such a neural network significantly outperforms parameter reconstruction from conventional high-level variables while at the same time providing reliable uncertainty estimates. 
Moreover, the neural network can quickly be re-trained for different detector properties, making it an ideal framework for optimizing experimental design. }

\begin{document}
\title{Reconstructing axion-like particles from beam dumps with simulation-based inference}

\author{Alessandro~Morandini \inst{1}\thanks{alessandro.morandini@kit.edu}, Torben~Ferber\inst{2}\thanks{torben.ferber@kit.edu}  and Felix~Kahlhoefer\inst{3}\thanks{kahlhoefer@kit.edu}}

\institute{Institute for Astroparticle Physics (IAP),  Karlsruhe Institute of Technology (KIT), D-76131 Karlsruhe, Germany \and Institute of Experimental Particle Physics (ETP),  Karlsruhe Institute of Technology (KIT), D-76131 Karlsruhe, Germany \and Institute for Theoretical Particle Physics (TTP),  Karlsruhe Institute of Technology (KIT), D-76131 Karlsruhe, Germany}

\authorrunning{A. Morandini, T. Ferber, F. Kahlhoefer}
\titlerunning{Reconstructing ALPs from beam dumps with simulation-based inference}






\maketitle

\section{Introduction}

The goal of any particle physics experiment is to gain insight into the underlying physical theory by using the recorded events to perform statistical inference. 
A common situation in high-energy physics is that one can easily simulate large numbers of events for given theory parameters, but there is no direct access to the likelihood of a given event. 
The resulting difficulty to infer theory parameters from observed events is called the \emph{inverse problem}.\footnote{For an introduction to the inverse problem and modern ways to address it see Refs. \cite{cranmer2020frontier,Brehmer:2020cvb}.} 
Its most common solution is to engineer a small number of high-level observables, whose probability distribution can be easily extracted from simulations. 

For events with a high multiplicity of final state particles, many different high-level observables can be constructed, and finding the optimal ones is an important and difficult task. 
But even a small number of final state particles may pose a challenge, if their properties are difficult to measure.
For example, consider the decay of a long-lived particle with unknown mass and lifetime into a pair of photons. 
Since photons do not leave tracks in the tracking detector, it is difficult to accurately measure the direction of their momentum in the electromagnetic calorimeter, which prevents an accurate reconstruction of the decay vertex and of the invariant mass of the parent particle~\cite{Alonso-Alvarez:2023wni}. 
In such a case different vertex reconstruction algorithms are needed, which construct complex high-level observables out of all the available experimental information that go beyond reconstructing four-vectors.

In such a setup, the optimal observable itself may depend on the details of the experiment, such as the size and position of the detector and its angle and energy resolution. 
In order to optimise experimental design, it then becomes necessary to automate the process of constructing high-level observables. 
Major progress has been made in this context in recent years by applying Machine Learning (ML) techniques to the physical sciences \cite{Carleo:2019ptp} and more specifically to high-energy physics \cite{Albertsson:2018maf}. Of particular importance for us is the application of ML to LHC physics \cite{Larkoski:2017jix,Guest:2018yhq} and to searches for new physics \cite{Karagiorgi:2021ngt}. The theory behind using ML to learn new physics has been studied in detail in Refs.~\cite{DAgnolo:2018cun,dAgnolo:2021aun,DeSimone:2018efk}.\footnote{For a complete list of works in particle physics making use of ML we recommend the living review \cite{Feickert:2021ajf}.}

ML approaches to the inverse problem have the key advantage that they are able to adapt to different detector designs easily, as long as the final state under consideration and the underlying physical process remain the same. For example, if we want to assess the impact of varying the calorimeter resolution on our ability to constrain the parameters of a specific model, it is typically enough to just retrain a neural network (NN) developed for a specific experimental setup without changing the underlying network architecture or training strategy.

An application of particular interest for this problem is the proposed construction of new beam-dump experiments at CERN to search for feebly-interacting particles at the GeV scale with macroscopic decay lengths. 
Among the most well-motivated such particles is an axion-like particle (ALP), which arises as a nearly massless particle from the spontaneous breaking of an approximate global symmetry~\cite{Dolan:2017osp}. 
If these ALPs couple dominantly to the electroweak gauge bosons of the SM, they may be copiously produced in rare meson decays (such as $B \to K + a$) and subsequently decay into pairs of photons. 
There already exist many constraints on such a scenario, but near-future experiments such as SHADOWS~\cite{Alviggi:2839484}, SHiP~\cite{SHiP:2021nfo} or HIKE~\cite{HIKE:2022qra} offer unique opportunities to substantially improve sensitivity.

In this work we explore a ML approach known as \emph{simulation-based inference (SBI)} in order to obtain likelihoods (or posterior probabilities) and reconstruct the ALP parameters from a small number of observed events.
We find that this approach adapts easily to variations in the assumed detector properties: If the properties of the final-state particles can be accurately measured, the SBI performs very similar to conventional methods that would reconstruct the vertex position and the invariant mass. 
If, on the other hand, only less accurate measurements are available, the network makes use of additional and correlated information, such as the angular and energy distribution of ALPs produced in rare meson decays, to significantly outperform conventional methods. 
Most importantly, this process is fully automated and can be quickly repeated for different detector designs, making it possible for example to perform cost-benefit analyses for a large number of experimental concepts.

The remainder of this work is structured as follows. 
In section~\ref{sec:simulator} we introduce the physics model that we consider, the typical experimental setup and how to simulate the physical processes to obtain mock data. 
In section~\ref{sec:SBI} we discuss possible ML approaches to analyse this data and identify conditional invertible neural networks as particularly promising. 
We then consider ALP parameter inference for different experimental designs in section~\ref{sec:applications} and discuss our results in section~\ref{sec:results}.

\section{ALPs at beamdumps}\label{sec:simulator}
\subsection{Event generation} 

ALPs can be generated and detected in different ways, depending on the underlying model parameters and the experimental design~\cite{Jerhot:2022chi}. 
In this work we focus on ALPs that are produced in the decay $B \to K + a$ and subsequently decay into photon pairs. 
Such a scenario arises for example from ALPs that couple dominantly to $SU(2)_L$ gauge bosons~\cite{Izaguirre:2016dfi}:
\begin{equation}
    \mathcal{L} \supset - \frac{g_{aW}}{4} a W^{\mu\nu} \tilde{W}_{\mu\nu} \; ,
\end{equation}
where $W^{\mu\nu}$ denotes the $SU(2)_L$ field strength tensor and $\tilde{W}^{\mu\nu}$ its dual. 
Alternatively, such a scenario can result from ALPs coupled to SM fermions as long as the decay into photons dominates (which requires $m_a < 3 m_\pi$ and suppressed couplings to SM leptons)~\cite{Dolan:2014ska,Carmona:2021seb,DallaValleGarcia:2023xhh}.

In our study, we will focus on a simple experimental setup inspired by NA62~\cite{NA62:2017rwk} and its proposed successors HIKE~\cite{HIKE:2022qra}, SHiP~\cite{SHiP:2021nfo} and SHADOWS~\cite{Alviggi:2839484}. 
These experiments are proposed to be placed in the ECN3 experimental hall at CERN after the SPS accelerator stage at the LHC. 
The extracted protons have an energy of $\SI{400}{\giga\electronvolt}$, which is enough to produce bottom and charm mesons when impinging on a fixed target.
The corresponding production cross sections and differential distributions have been studied in Ref.~\cite{Dobrich:2018jyi,Afik:2023mhj}, and we take their PYTHIA8~\cite{10.21468/SciPostPhysCodeb.8} samples of $B$ mesons as starting point for our simulations. 
ALP production through $D$ meson decays is found to be negligible~\cite{Jerhot:2022chi}, so we focus on $B$ meson decays instead.

The decay of ALPs into photon pairs is described by the Lagrangian
\begin{equation}
\mathcal{L}\supset -\frac{1}{4}g_{a\gamma}aF_{\mu\nu}\tilde{F}^{\mu\nu} \; ,
\end{equation}
where the effective ALP-photon coupling is related to the coupling to $SU(2)_L$ gauge bosons via $g_{a\gamma} =   \sin^2 \theta_W g_{aW}$ with $\theta_W$ being the weak mixing angle. 
The ALP lifetime is given in terms of this coupling and the ALP mass $m_a$ by the relation
\begin{equation}
(c\tau_a)^{-1}=\Gamma_{a\rightarrow\gamma\gamma}=\frac{g_{a\gamma}^2}{64\pi}m_a^3.
\end{equation}
We will be interested in the case where this coupling is small (of the order of $10^{-5} \, \mathrm{GeV}^{-1}$), which implies a macroscopic decay length up to hundreds of meters.

Both the $B$ meson decay and the subsequent ALP decay are isotropic in the respective rest frames, such that the distribution of photon angles and energies can be easily obtained through appropriate rotations and Lorentz boosts. We simulate these decays following the public code ALPINIST~\cite{Jerhot:2022chi}.
To determine the position of the ALP decay vertex, we assume that the $B$ meson decays promptly at $\mathbf{r} = (0,0,0)$ and sample randomly from the exponential distribution of ALP decay lengths $d$ given by
\begin{equation}\label{eq:decay_exp}
    p(d) = \exp\left( - \frac{d \, m_a}{p_a c \tau_a}\right) \; ,
\end{equation}
where $p_a$ denotes the ALP momentum in the laboratory frame. The vertex position is then obtained as $\mathbf{r}_V \equiv (x_V, y_V, z_V)= d \, \mathbf{p}_a / p_a$.\footnote{We note that, while the vertex distribution is invariant under rotations in the plane transverse to the beam direction, most detector geometries are not. To maintain generality, we will therefore retain all three spatial coordinates.}

In principle, the branching ratios for $B \to K + a$ can also be calculated in terms of the effective ALP model parameters. 
In this work, however, we will treat the $B$ meson branching ratios, and hence the ALP production cross section, as an independent parameter. 
This is well-motivated both in the case of gauge boson interactions, where the effective ALP photon coupling may receive an additional contribution from an underlying ALP-hypercharge coupling, and in the case of quark interactions, where the $B$ meson branching ratio has a residual logarithmic dependence on the ultraviolet completion~\cite{Dolan:2014ska,Alonso-Alvarez:2021ett}. 
The $B$ meson branching ratios then only affect the total number of expected events, i.e.\ the normalisation of the various distributions, but not their shape. 
In the following, we will focus our attention primarily on the two ALP parameters that affect kinematic distributions in more complicated ways, namely $(m_a, g_{a\gamma})$ or equivalently $(m_a, c\tau_a)$. These are provided as input to our simulator in order to extract experimental observables.

\subsection{Detector geometry and experimental setup}

\begin{figure}[t!]
\centering
\includegraphics[width=0.5\textwidth]{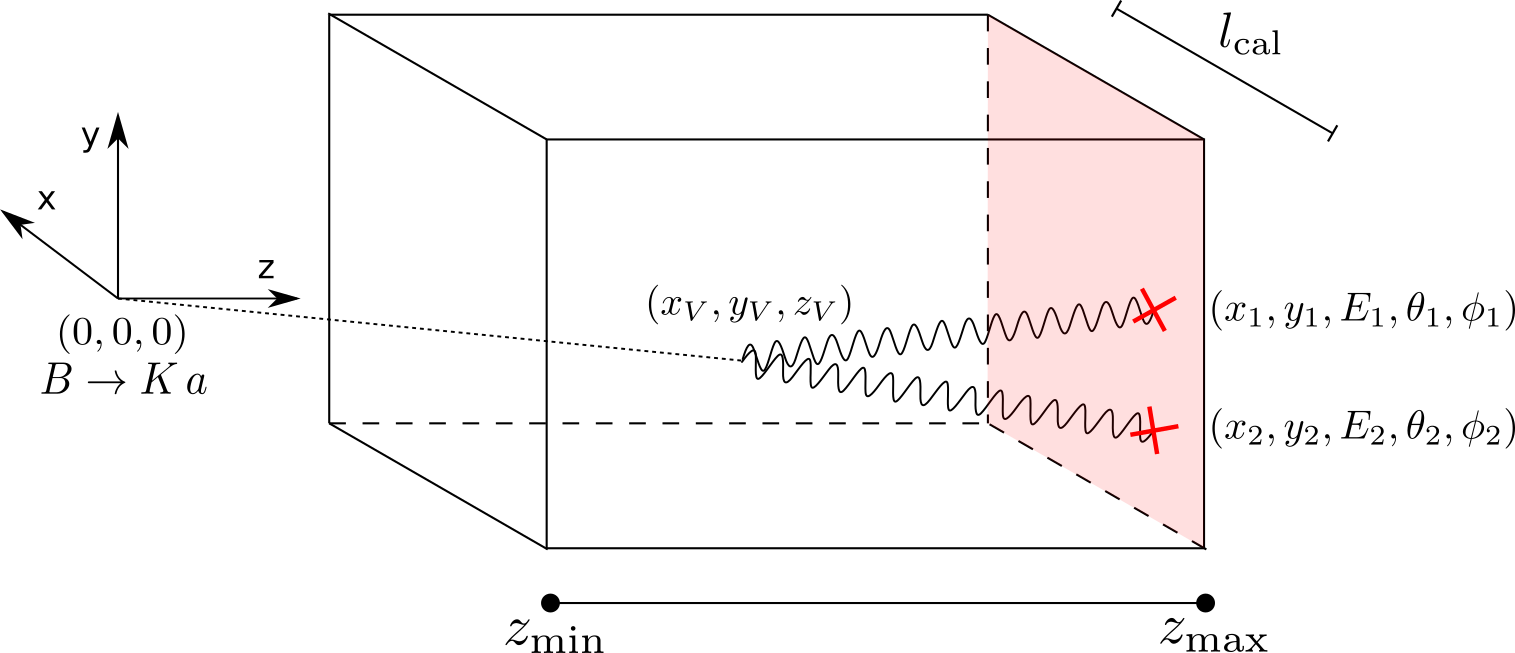}
\caption{Sketch of the detector design, with a focus on the observable features ($x_i,y_i, E_i,\theta_i, \phi_i$). The calorimeter plane has been highlighted in light red.}
\label{fig:detectorsketch}
\end{figure}

We consider a typical beam-dump experiment, where the ALPs are produced inside an absorber and propagate into an evacuated decay volume~(see Fig.~\ref{fig:detectorsketch}). 
The photons produced in the ALP decays then propagate through the decay volume and are detected when interacting with the calorimeter at the far end of the experiment. 
The decay volume is placed at a distance $z_\text{min}$ from the point where the proton beam impinges onto the dump. 
The decay volume ends at $z_\text{max}$ where the calorimeter observing the photons is located. 
The calorimeter is assumed to be a square with side length $\ell_\text{cal}$ centred at $x = x_\text{cal}$ and $y = 0$. 
In a more refined treatment, we would need to take into account that between the end of the decay volume and the calorimeter there are tracking detectors. 
In our simplified discussion the tracking chambers are taken to be part of the decay volume and thus the calorimeter is placed directly at the end of the decay volume. 

Candidate ALP events are selected if both photons hit the calorimeter plane. In order to ensure that the resulting showers can be individually resolved, we require a minimum photon separation of $d_\text{min}=\SI{10}{\centi\meter}$. Furthermore, we require both photons to have an energy greater than $\SI{1}{\giga\electronvolt}$, which is readily satisfied by photons produced in the decay of a boosted ALP.
A perfect detector would be able to reconstruct the photon 4-momenta (i.e.\ their energy $E_i$ and angular information $\theta_i, \phi_i$) and the calorimeter hit position ($x_i$ and $y_i$). 
Experimentally, the showers position $z_i$ needs to be determined as well.
Since the shower $z$-coordinate is typically meters away from the decay vertex, we identify $z_i$ with the position $z_\text{max}$ of the first calorimeter plane and assume that the impact of the uncertainty is small compared to the other uncertainties.
We note that these observables contain redundant information: The requirement that both photons originate from the same decay imposes one constraint on the ten observables ($E_i$, $\theta_i$, $\phi_i$, $x_i$ and $y_i$), while two further constraints are obtained in the case of a two-body decay, even if the mass of the decaying particle is unknown. 
If the observables are consistent with these constraints, it is possible to reconstruct the vertex position ($x_V, y_V, z_V$).

Typical laterally segmented electromagnetic calorimeters provide relative photon energy resolution of a few percent for GeV-energies.
The shower position can be reconstructed to a fraction of the spatial segmentation.
These six observables $(E_i, x_i \text{ and } y_i)$ are generally insufficient to reconstruct the vertex position or the ALP mass.
To do so, we need to extract at least some amount of angular information from the electromagnetic showers, such as the photon opening angle
\begin{equation}
    \alpha_{\gamma \gamma} = \text{arccos}\left(\frac{\mathbf{p}_1 \cdot \mathbf{p}_2}{p_1 \, p_2} \right) \; .
\end{equation}
The accuracy with which $\theta_i$ and $\phi_i$ (and hence $\alpha_{\gamma\gamma}$) can be measured will directly affect our ability to reconstruct the underlying physical process.
The measurement uncertainty critically depend on the detector properties, as for instance the cell granularity and absorber material. To fully characterize the experimental setup, we therefore need to define the accuracy with which we can measure the different quantities in addition to specifying the detector geometry.

We will consider two different detector geometries in our study, which we call ``on-axis detector'' and ``off-axis detector''. The size of the decay volume and the calorimeter is the same in both cases. We have $z_\text{min}=\SI{10}{\meter}$, $z_\text{max}=\SI{35}{\meter}$ with a calorimeter size $l_\text{cal}=\SI{2.5}{\meter}$. The two detector geometries differ in the fact that for the on-axis case $x_\text{cal}=0$ and for the off-axis case $x_\text{cal}=\SI{2.25}{\meter}$. Our off-axis geometry has been inspired by the current SHADOWS proposal. A detector with fixed geometry will then be characterized by a set of three uncertainties: the angular resolution (which we approximate to be the same on the polar angle $\sigma(\theta)$ and on the azimuthal angle $\sigma(\phi)$), the relative energy uncertainty $\sigma(E)/E$ and finally the calorimeter hit resolution $\sigma(h)$.

To conclude this section, we emphasize that, while there is redundant information in the ten observables measured by the experiment, it is necessary to use all of them to extract as much information as possible. Traditional approaches to this problem employ vertex reconstruction algorithms based on the photon angles measurement, such that the reconstruction error of the decay vertex depends on the accuracy with which we can measure the photon momenta and calorimeter hit positions. 
The algorithms that we introduce below, are not explicitly required to extract vertex information, but they may of course learn such information if possible and necessary. 

\begin{figure*}
    \centering
    \includegraphics[width=0.9\textwidth]{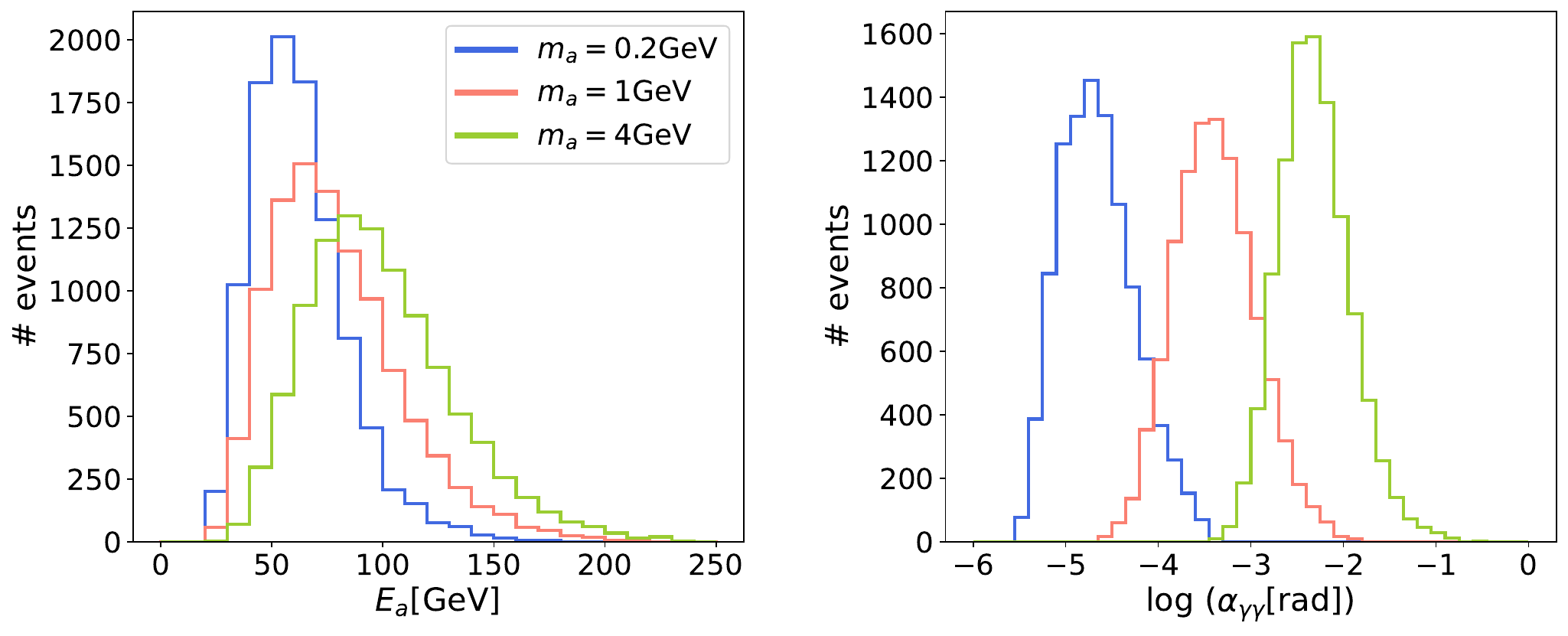}
    \caption{Distribution of generated quantities for varying ALP masses. The energy of the ALP can be reconstructed as the sum of the single photon energies. The photon angular separation requires the measurement of the photon momenta.}
    \label{fig:distributions}
\end{figure*}

We emphasize, however, that an accurate vertex reconstruction is not always necessary to address the inverse problem. If we look at \cref{fig:distributions}, we can see that even something as low-level as the ALP energy, which can be reconstructed as the sum of the photon energies, contains information about the ALP mass, because the ALP energy distribution is determined by the underlying ALP production mechanism through $B$ meson decays. 
However, the ALP energy distributions differ only slightly for different masses, so unless enough statistics is provided we will not be able to put a strong constraint on the ALP mass from this information alone. 
The opening angle between photons, on the other hand, is harder to measure precisely, but the distribution of this observable is very informative about the ALP mass.

\section{Simulation-based inference}

\label{sec:SBI}

\subsection{Motivation}

As discussed above, it is straightforward to generate events for a given set of ALP model parameters. 
The inverse problem of inferring ALP parameters from one or more observed events is generally much harder. 
A promising strategy could be to try and reconstruct the decay vertex and the invariant mass of the decaying particle, from the observed photons. 
This is easily possible if the position and momentum of each photon is known with high accuracy.

In practice, sizeable measurement uncertainties prevent an accurate reconstruction of the vertex and invariant mass. 
While it may still be possible to estimate the vertex position and invariant mass using for example a neural network~(NN) regressor, the statistical interpretation of the output is unclear. Even if the regressor is trained to predict the uncertainty of its estimate, or if the uncertainty is inferred from simulations, this information would typically only be useful if the deviations follow approximately a normal distribution.\footnote{The possibility of suitably modifying the loss function of a NN regressor to not only provide an estimator, but also an uncertainty on the estimator has been considered in the literature \cite{Jeffrey:2020itg,CAMELS:2021raw}. 
This latest approach could be an alternative in the case approximating the posterior is not feasible, but when it is possible our approach leads to better control due to its well-defined statistical foundation.}

For a rigorous statistical inference it is indispensable to know the likelihood function, regardless of whether one uses frequentist or Bayesian methods \cite{Zech:2001eh,bohm2010introduction}.
Even if the likelihood is intractable analytically, it can be reconstructed using an approach called simulation-based inference. 
As the name suggests, the main ingredient of this approach are simulations, i.e.\ samples of events drawn from the likelihood. 
While the approach does not in principle require ML, it has greatly benefited from advancements in ML algorithms, which enabled its application in high-dimensional problems and has lead to growing popularity in recent years~\cite{rezende2016variational,papamakarios2021normalizing,cranmer_kyle_2016_198541,Cranmer:2015bka,Baldi:2016fzo}. 

A key advantage of simulation-based inference is that it uses and combines all information available in the observed events. 
For example, since we assume a specific ALP production process, the energy of the ALP which can be inferred from the energies of the two photons even if the decay vertex cannot be reconstructed is correlated with the ALP mass (see \cref{fig:distributions}). 
If additional information on the invariant mass or the decay vertex are available, it will be automatically combined with other kinematic variables. 
Since there is no need to construct explicit high-level observables, no information is lost and the most accurate estimates are obtained. 
The remaining uncertainties can be directly extracted from the shape of the likelihood (or posterior). 
This makes it comparably easy to separate the irreducible physical uncertainty due to inaccurate measurements from the reducible network uncertainty.

Common ML algorithms for simulation-based inference include classifiers \cite{Miller:2020hua,Miller:2022shs}, which learn probability ratios, and normalising flows (NFs) \cite{dinh2017density,NEURIPS2018_d139db6a}, which perform a density estimation task by minimizing deviations between the predicted probability distribution and the one of a given sample. While classifiers do not learn the distributions directly, they have the key advantage of being trained on an easier learning task. This implies that they are in principle better suited to tackle hard problems. In practice, however, we have found that simple classifiers often struggle to reproduce very narrow posteriors. Such posteriors are expected in our set-up for the case of excellent detector resolution.
In the following we will thus focus on the NF algorithm and consider a specific modification called conditional invertible neural networks~(cINN) \cite{ardizzone2019guided,BayesFlow}, described in detail below. NFs are commonly harder to train for difficult tasks, but in our case the posterior is unimodal and low-dimensional. We have observed good convergence and stability of the training, confirming that the cINN is an appropriate algorithm to deal with the task at hand.

We note that simulation-based inference can be applied both in the frequentist approach (in order to obtain likelihoods or likelihood ratios) and in the Bayesian approach (where one focuses on posteriors or likelihood-to-evidence ratios). 
While the ML algorithm discussed below can be adapted to either case, we will focus on the Bayesian approach, as it is more intuitive given the low dimensionality of our parameter space and the high variability of the observations. Since the cINN has learnt the posterior, we are able to sample directly from it. This has some key practical advantages, as we can quickly derive the marginal posteriors and the credibility regions once we have the posterior samples.
For a frequentist approach, the network would need to be adapted suitably \cite{ardizzone2019guided}. In this case, with the cINN we would sample from the likelihood, providing an alternative event generator.

\subsection{Normalizing flows and conditional invertible neural networks}

Normalizing flows tackle the issue of density estimation by transforming a known probability density function~(PDF) through a suitable change of variables. 
To make this more formal, let us consider a random variable $\mathbf{z}$ distributed under a known PDF $f(\mathbf{z})$. 
If a new random variable $\mathbf{x}$ is defined through a bijective transformation $\mathbf{z}=g(\mathbf{x})$, its PDF is given by
\begin{equation}
p(\mathbf{x}) = f(g(\mathbf{x}))|\det J|,\quad\text{with } J_{ij} =\frac{\mathrm{d}\mathbf{z}}{\mathrm{d}\mathbf{x}}=\frac{\partial z_i}{\partial x_j}.
\end{equation}
In order to estimate the PDF of a given sample $\mathbf{x}$, we choose a convenient form for $f(\mathbf{z})$, for example a Gaussian distribution and give a NN the task of finding a suitable transformation $g(\mathbf{x})$. 
Such a NN will be defined by its architecture and  a set of weights $\psi$, so that we can write $g_\psi(\mathbf{x})$ to denote the family of transformations $g$ which can be expressed by the NN for varying weights $\psi$. 
The optimal transformation will be the one that maximizes the probability of the sample, or equivalently minimizes the Kullback-Leibler divergence~(KL) between the true distribution $p$ and the proposal distribution $p_\psi$:
\begin{align}\label{eq:DKL}
D_\text{KL}(p||p_\psi) = -\frac{1}{N}\sum_i^N  \log p_\psi(\mathbf{x}_i)  \\ \text{with} \quad p_\psi(\mathbf{x}) \equiv f(g_\psi(\mathbf{x}))|\det J_\psi| \; .
\end{align}
We denote the function $p_\psi$ that minimizes $D_\text{KL}(p||p_\psi)$ by $\tilde p$. 
To make the NN more expressive, we actually stack several transformations $g_l$. The result is a series (flow) of transformations which maps our target distribution $p(\mathbf{x})$ to a normal distribution $f(\mathbf{z})$, hence the name \emph{normalizing flow}.

In the case at hand, we have a high dimensional vector $\mathbf{x}$ distributed according to a likelihood $\mathcal{L}$ that depends on model parameters $\boldsymbol{\theta}$. 
Lacking knowledge of the true values of $\boldsymbol{\theta}$, we can consider different random choices $\boldsymbol{\boldsymbol{\theta}}_i$ following an assumed prior probability $\pi(\boldsymbol{\theta})$. 
For each such choice, we can generate a random event $\mathbf{x}_i$ from $\mathcal{L}(\boldsymbol{\theta}_i)$. 
Given pairs of $\boldsymbol{\theta}_i$ and $\mathbf{x}_i$, NFs will learn the joint distribution
\begin{equation}
    p(\mathbf{x}, \boldsymbol{\theta})=\mathcal{L}(\mathbf{x}|\boldsymbol{\theta})\pi(\boldsymbol{\theta})=p(\boldsymbol{\theta}|\mathbf{x})p(\mathbf{x}).
\end{equation}
From this distribution both the posterior and the likelihood can be derived dividing by the evidence and the prior respectively. The NN structure can be adapted to guarantee that the function $g_\psi$ is invertible and the determinant of the Jacobian $J_\psi$ fast to compute \cite{NIPS2016_ddeebdee,dinh2017density}.

In the example above, the NFs treat in the same way the low-dimensional model parameters and the high-di\-mensional observables, which can lead to difficulties when learning the joint distribution. This problem is addressed by cINNs, which are able to directly learn conditional probabilities, i.e.\ likelihoods or posteriors. In previous works, cINNs have been employed in high-energy physics for event generation \cite{Butter:2020tvl,Butter:2021csz,Butter:2022rso,Kach:2022qnf}, unfolding \cite{Bellagente:2020piv,Backes:2022vmn} and anomaly detection \cite{Hallin:2021wme,Fanelli:2022xwl}, and for inference in other physical scenarios like the measurement of QCD splittings~\cite{Bieringer:2020tnw} and the study of cosmic rays~\cite{Bister:2021arb}.
Like NFs, cINNs work with density transformations, but now the model parameters and the observations enter the network differently. 
For example, we can transform the distribution of $\boldsymbol{\theta}$ into a normal distribution, while $\mathbf{x}$ is not transformed and is used to determine the transformation of $\boldsymbol{\theta}$.
Proceeding in this way we would determine the probability of $\boldsymbol{\theta}$ conditioned on $\mathbf{x}$, which is the posterior probability. 
Analogously, we can transform $\mathbf{x}$ and use $\boldsymbol{\theta}$ to determine the transformation, in which case we would derive the probability of data conditioned on the model parameters, i.e.\ the likelihood. 

\begin{figure}[t]
\centering
\includegraphics[width=0.45\textwidth]{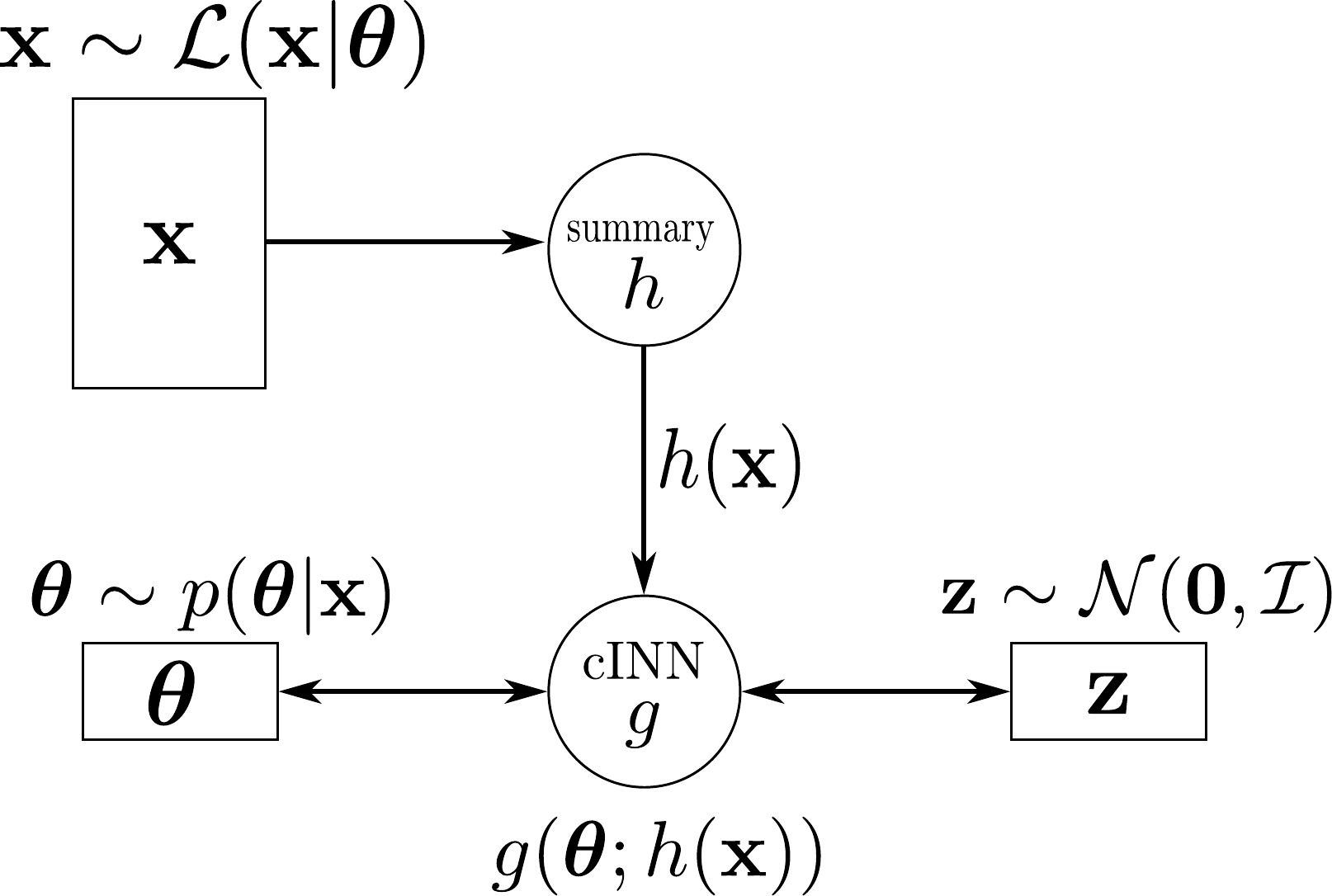}
\caption{\label{fig:cINN} Schematic view of our network architecture, with focus on the summary network component and the cINN. The summary network $h$ is a fully connected neural network, while $g$ is an invertible neural network with affine coupling layers. A full description of the architecture and hyperparameters is given in \cref{tab:architecture}.}
\end{figure}

While conditional invertible neural network treat observables and model parameters differently, the issue that the dimensionality of the observables is considerably larger than the dimensionality of the model parameters remains. One way to improve the stability of training is to use a summary network $h(\mathbf{x})$, which is trained in tandem with the cINN itself. This summary network takes as input the high-dimensional low-level observables and provides as output to the cINN a vector of lower dimension (see \cref{fig:cINN}). We then have a set of weights $\zeta$ defining the summary network and a set of weights $\psi$ defining the cINN transformation, but both of them are determined by the same training loop. In other words, we aim to minimize $D_\text{KL}(p||p_{\psi\zeta})$, where
\begin{equation}
 p_{\psi\zeta}(\boldsymbol{\theta}|\mathbf{x}) \equiv f(g_\psi(\boldsymbol{\theta}; h_\zeta(\mathbf{x})))|\det J_{\psi\zeta}| \; . 
\end{equation}
By training the summary network at the same time as the cINN, we effectively ask the network to learn the high-level observables that are best suited to constrain the mass and the lifetime. The final result should not be qualitatively different provided the mutual information between the input variables and the learnt high-level observables is large \cite{BayesFlow}. For the scenario under consideration we have observed that using a summary network leads to a considerably more stable training. Not only that, but the use of a summary network simplifies the network learning and can even lead to better performance.
Clearly, the learnt high-level observables may not be the same for different detector setups. 

To conclude this discussion, we remind the reader that the discrete estimator of the KL divergence in \cref{eq:DKL} is defined up to a constant that depends solely on the true posterior. The network can thus be trained directly on the task of minimizing the distance from the true posterior without knowing it explicitly. However, the lack of knowledge of the true posterior implies that one cannot immediately quantify whether the posterior approximation is accurate. In the following, we will therefore explore alternative strategies to ensure that the posteriors obtained from the cINN are sufficiently close to the true posterior.

\subsection{Application to ALPs}

Let us now discuss how the general discussion above applies to our training samples and variables. 
First of all, we need to specify the probability distributions for the model parameters, which are used to generate the training samples. In our Bayesian approach, these are also the prior probabilities. Since we are interested in ALPs produced in rare $B$ meson decays, we focus on the mass range where ALPs would be too heavy to be produced in $K \to \pi + a$ thus evading strong constraints from Kaon experiments like NA62 but light enough to be produced in $B \to K + a$. 
We therefore sample the masses with a log-prior on $[\SI{0.1}{\giga\electronvolt}, \SI{4.5}{\giga\electronvolt}]$.\footnote{In principle the mass range under consideration is small enough that a uniform prior could also work, but in order for our approach to be easily generalisable, the log-prior is more appropriate.} 
The detector under consideration has $z_\text{min}=\SI{10}{\meter}$, $z_\text{max}=\SI{35}{\meter}$, and only particles with boosted lifetime comparable to these dimensions can be efficiently detected. 
We therefore consider proper lifetimes in $[\SI{0.1}{\meter}, \SI{100}{\meter}]$ with a log-prior. 
We then construct the input parameter vector given by
\begin{equation}
    \boldsymbol{\theta} = (\log_{10}(m_a[\SI{}{\giga\electronvolt}]), \log_{10}(c\tau_a/m_a[\SI{}{\meter}/\SI{}{\giga\electronvolt}])) \; . 
\end{equation}
We note that the unboosted lifetime $c\tau_a/m_a$ is what enters \cref{eq:decay_exp}. 
Considering this combination (rather than the lifetime $c \tau_a$) reduces degeneracies and hence decorrelates the two model parameters, at least if the calorimeter has sufficient resolution to reconstruct the vertex position.

When the cINN is trained, it is only implicitly aware of the priors, as they affect the distribution of model parameters given to the network for training. 
This is different to the case of a classifier used to learn the likelihood-to-evidence ratio, where the prior needs to be given explicitly~\cite{Miller:2022shs}. 
The fact that the prior is given implicitly means that the posterior estimated by the cINN does not need to be identically zero outside of the prior range. 
When evaluating the posterior for parameter points close to prior borders, we see a smooth transition to zero, rather than a step. 
It would be possible to truncate and normalize the posterior to force it to be zero outside of the prior range, if there is a physical reason to do so. 
In our case the prior ranges are set by defining the regions where we expect the experiment to have sufficient sensitivity, but there is no physical reason why the lower or higher lifetimes should not be at all possible. 
This does not apply to the upper value on the mass, as for our production process $B \to K + a$ the ALPs cannot have a mass larger than $m_B-m_K\approx\SI{4.78}{\giga\electronvolt}$, so the posterior for larger masses should be identically zero if $B$~meson decays are the only relevant ALP production process. 
In our case we settle for a less stringent upper bound on the mass of $\SI{4.5}{\giga\electronvolt}$, so that we do not encounter this physical upper limit.

Having specified how we sample the model parameters, let us move to the event observables. 
As discussed above, each ALP event is characterised by ten experimental observables, namely the two photon energies, their the 2D-shower positions and the shower directions. 
Even though our setup in principle works with just one observed event, we will consider data sets consisting of three observed events for each pair ($m_a, c\tau_a/m_a$) of model parameters. 
Our observable space therefore has dimensionality $D=30$. 
The three events are not ordered in any way, while the two photons are ordered based on their energy. 
To improve the training, before passing the observables to the network, we take the natural logarithm of the photon energies and the photon polar angles to avoid inputs that can vary over orders of magnitude. 
Finally, both our model parameters and observables receive standard preprocessing so that they have zero mean and unit variance. 
For our analysis we have decided to consider sets of three observed events, but our discussion applies in much the same way to any number of observed events. 
However, it is useful to work with multiple observed events, because we can then perform consistency checks between them and reduce the risk of background contamination. 
Our specific choice is motivated by the common use of three predicted events for the sensitivity projection in background-free experiments. 
Let us clarify our naming conventions: we will call a single diphoton measurement an event, we combine three events into (data) sets and our training/test samples will consist of multiple sets.

For the specifics of the network architecture and training, our setup is based on Ref.~\cite{BayesFlow}, but adapted for the task at hand. 
We fix the output of the summary network to be two-dimensional in order to obtain two high-level observables that are informative of the mass $m_a$ and the unboosted lifetime $c\tau_a/m_a$, respectively. 
We have checked that increasing the output dimension of the summary network to three or four does not lead to qualitatively different results. 
We have optimized the other hyperparameters by performing a scan over them for a fixed detector setup. More precisely, we have considered arrays of possible hyperparameter values and combined them to have multiple network trainings. Among all the possibilities we have looked at the six with the lowest validation losses. We have then applied these six combinations of hyperparameters to the other detector setups to confirm that also for them we would get good training performances. In the end we picked the hyperparameter combination that lead to the lowest validation loss. A summary of the network architecture is given in \cref{tab:architecture}. 

For all the cases that we will consider below, we find that the architecture and training hyperparameters given in \cref{tab:architecture} yield good convergence of the training and validation loss. Thanks to early stopping we avoid overfitting due to over-training, typically for our detector setups we see a difference between the validation and training loss of $5\%$ with respect to the total training loss improvement from the first epoch to the end of training. We will explicitly show below that the residual overfitting, while not completely negligible, does not introduce a significant bias in our results.

\begin{table}
    \centering
    \begin{tabular}{c|c}
         Coupling layers &  \\
         \hline
         Number coupling layers & 4 \\
         Hidden layers & [128, 128, 128, 128] \\
         Hidden layers activation & ReLU \\
         Output layer activation & linear \\
         \hline \hline
         Summary network &  \\
         \hline
         Output layer dimension & 2 \\
         Hidden layers & [64, 64, 64, 64] \\
         Hidden layers activation & LeakyReLU ($\alpha = 0.01$) \\
         Output layer activation & linear \\
         \hline \hline
         Training hyperparameters &  \\
         \hline
         Max number epochs & 500 \\
         Batch size & 512 \\
         Initial learning rate & $5\cdot 10^{-3}$ \\
         Decay rate & 0.9 every 10 epochs \\
         Early stopping & $\delta < 10^{-3}$ for 50 epochs\\
    \end{tabular}
    \caption{Architecture of the summary network and of the cINN. The output of the summary network is fed into the coupling layers transformations. Since the summary network and the cINN are trained together, the training hyperparameters apply to both of them.}
    \label{tab:architecture}
\end{table}

\section{Applications}
\label{sec:applications}

In this section we apply the cINN introduced above to the production and detection of ALPs in proton beam dumps, considering a simplified experiment.
Before assessing the performance of different detector setups, we take a closer look at the posterior in order to understand what is easy and hard for the network to learn.
We also discuss how to deal with background events and the effect of changing the detector resolution for low-level observables. 

\subsection{Learning the posterior}
\label{sec:learningposterior}
In the following, we will measure the performance of a given detector setup by determining the width of the posterior, which tells us how tightly the underlying parameters can be constrained.
It is therefore instructive to visualize the posterior for some representative scenarios. 
Let us start with the detector geometry defined in section~\ref{sec:simulator}, in particular for now our detector is on-axis ($x_\text{cal}=0$). 
To understand what a typical posterior will look like, we consider a specific detector setup with $\sigma(E)/E = 0.05,\,\sigma(h)=\SI{0.1}{\centi\meter},\,\sigma(\theta)=\sigma(\phi)= \SI{5}{\milli\radian}$.
Such resolutions are achievable with the calorimeters proposed for the next generation of beam dump experiments.
We note that realistically energy and angular resolutions are function of the photon energy itself, with resolutions improving for increasing photon energies.

\begin{figure*}[t]
    \centering
    \includegraphics[width=0.45\textwidth]{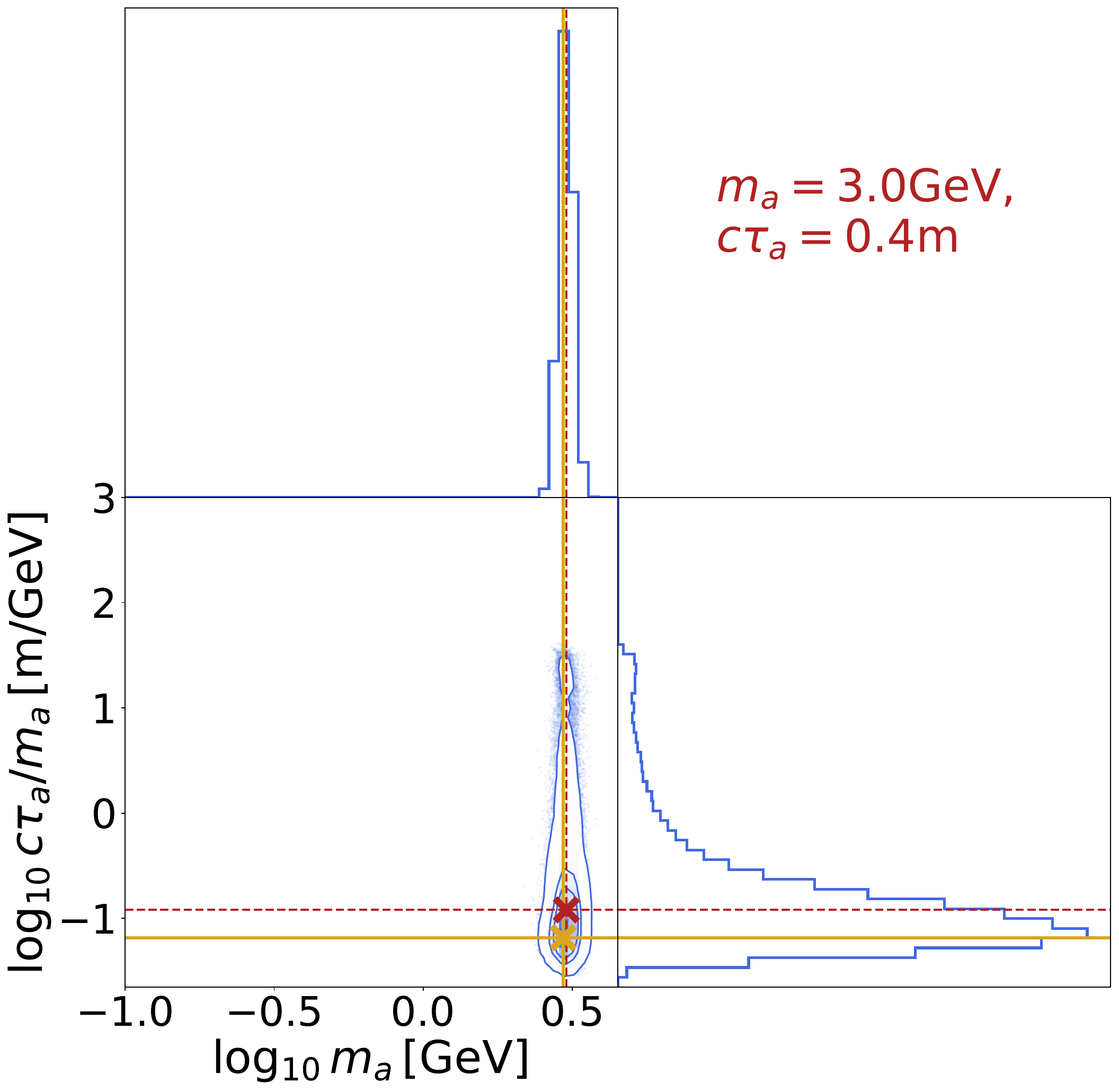}
    \quad\includegraphics[width=0.45\textwidth]{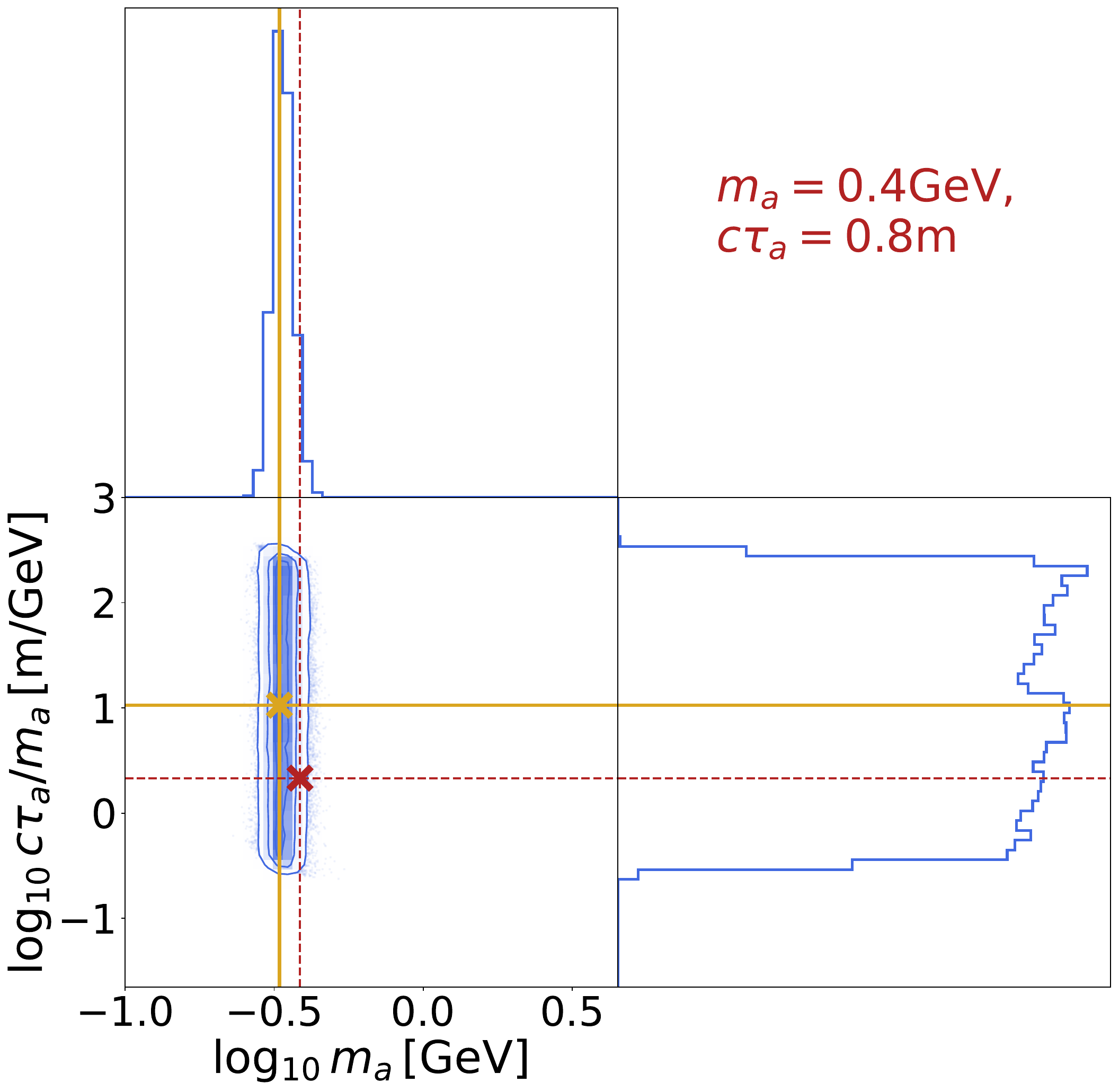}
    \caption{Joint posterior and marginal posteriors for two different observations. In red we have the true parameter values, in yellow we have the best fit to the joint posterior. The three contours indicate the $50\%$, $68\%$ and $95\%$ credible regions. The sampled posterior points outside of the $95\%$ credible region are also shown in the plot.}
    \label{fig:example}
\end{figure*}

First of all, we emphasize that the posterior for events generated by different assumed ALP masses and lifetimes will not have the same shape, in particular their spread will vary significantly. 
This is to say that physically not all lifetimes and masses are equally easy to reconstruct. As a general rule for our production mode and detector geometry, higher masses and lower (unboosted) lifetimes (within the sensitivity of the detector) are easier to constrain. 
We show the reconstructed posterior for two different observations in \cref{fig:example} corresponding to two different model parameters. 
On the left, we have an ``easy'' to constrain observation, with a small lifetime and sizable mass. This parameter point leads to pretty distinctive signatures in the detector, like a decay position close to $z_\text{min}$ and large opening angles $\alpha_{\gamma\gamma}$. On the right, we have an ``hard'' to constrain observation, corresponding to a smaller mass and a longer lifetime. In this case the marginal posterior on the mass is broader and the marginal posterior on the lifetime is no longer peaked but rather flat. 

In both cases we find that, as anticipated, the two parameters are largely uncorrelated, i.e.\ there are no non-trivial degeneracies in the posterior. 
The broad posterior for the unboosted lifetime is not due to a deficiency of the cINN, but reflects the fact that constraining the lifetime is fundamentally harder than constraining the ALP mass. 
This is easily understood, given that in the case of a perfect detector we would be able to precisely measure the ALP mass by constructing the diphoton invariant mass from a single event.
Contrary to this, the decay position stems from a (truncated) exponential distribution, meaning that even if we precisely measured the decay position, we would not be able to infer the lifetime exactly.

To make these statements more quantitative, we need to quantify the width of the posterior. 
Since the  posterior can be highly non-Gaussian, simple width estimators like the full width at half maximum are not descriptive of the posterior and cannot be used to compare different setups. 
However, the trained cINN enables us to directly sample from the posterior. 
It is then straightforward to evaluate the covariance matrix $\Sigma_{m, c\tau/m}$: 
\begin{equation}
\Sigma_{m, c\tau/m}=
    \begin{pmatrix}
\sigma_m^2 & \sigma_{m\,c\tau/m} \\
 \sigma_{m\,c\tau/m}  & \sigma_{c\tau/m}^2
\end{pmatrix}\;, \label{eq:covariance}
\end{equation}
which corresponds to the inverse of the Fisher information matrix~\cite{Edwards:2017mnf}. From this matrix we can determine the typical area of parameter space enclosed by the posterior as
\begin{equation}
    A_{m, c\tau/m}=\pi \sqrt{\det\left(\Sigma_{m, c\tau/m}\right)} \; .  \label{eq:area}
\end{equation}
In these equations and in the following we will write for brevity $\sigma_m$ and $\sigma_{c\tau/m}$, but these quantities refer to $\log_{10}(m)$ and $\log_{10}(c\tau/m)$ as these are our input parameters. 
The related uncertainties on the linear quantities can be derived via error propagation. 

While the area in eq.~\eqref{eq:area} is a good performance measure in general, in our case it is possible to make further simplifications. 
First, the off-diagonal terms in the covariance matrix are by construction small compared to the diagonal ones. 
Second, constraining the mass is generally easier than constraining the (unboosted) lifetime. This implies that best way to make the posterior narrow is by reducing the uncertainty in the mass, rather than in the lifetime. 
In our study we have found that while $\sigma_m$ can vary by up to two orders of magnitude, $\sigma_{c\tau/m}$ shows little variation and is largely determined by the geometry of the decay volume. The best way to improve the reconstruction of the lifetime would be to increase the length of the decay volume and place it closer to the interaction point. We have checked this for a detector geometry with $[z_\text{min}, z_\text{max}]=[\SI{2}{\meter}, \SI{100}{\meter}]$ and found that the posterior on the lifetime becomes more narrow and more peaked. 

For simplicity, and to allow for a more intuitive interpretation of our results, we will use $\sigma_m$ instead of $A_{m, c\tau/m}$ as performance measure in the following. 

It is clear from \cref{fig:example} that different ALP parameter points will generally have different posterior widths. 
For the ``easy'' example (left panel) we obtain $\sigma_m = \SI{0.024}{\giga\electronvolt}$, while the ``hard'' example (right panel) gives $\sigma_m = \SI{0.034}{\giga\electronvolt}$.\footnote{We note that the achievable mass resolutions is many orders of magnitude larger than the intrinsic width of the ALP corresponding to the assumed decay length.} The width of the posterior will depend on the model parameters used to generate the observed sample. 
It is therefore useful to keep these parameters fixed to some benchmark cases when comparing the performance of different detector setups. 
In this way we reduce the variability intrinsic to different regions of the parameter space. 
But even in the case we consider a fixed parameter point, we can have a large variance in the values of $\sigma_m$ obtained. In the following we will consider test data sets of 10000 samples and evaluate the distribution of $\sigma_m$ over these data sets. 

Before doing so, we however need to ensure that the confidence regions derived from the posterior are reliable. 
In other words, we need to verify that the posterior width is indicative of the uncertainty on the parameters, in particular the mass. 
We do not have access to the true posterior, so we cannot quantify the goodness of our approximation by direct comparison.
However, we can check the coverage \cite{dalmasso2021likelihood,hermans2021trust,Cole:2021gwr}.

Given a credibility level $\alpha$ and a posterior $p(\boldsymbol{\theta}|\mathbf{x})$, this defines a highest-posterior credible region where $p(\boldsymbol{\theta}|\mathbf{x})>p_\alpha$. Here, $p_\alpha$ is defined implicitly by requiring
\begin{equation}
\alpha = \int_{ p(\boldsymbol{\theta}|\mathbf{x})>p_\alpha} p(\boldsymbol{\theta}|\mathbf{x})\,\mathrm{d}\boldsymbol{\theta} \; .
\end{equation}
If we generate random sets $\mathbf{x}_j$ and evaluate our posterior on them, we would then expect that the true value $\bar{\boldsymbol{\theta}}$ lies in the credible region defined by credibility level $\alpha$ for a fraction $\alpha$ of events. 
This is our expected coverage. 
In our case we only have access to an approximation of the posterior and the coverage obtained from this approximation takes the name of empirical coverage. 

\begin{figure*}
    \centering
    \includegraphics[width=0.9\textwidth]{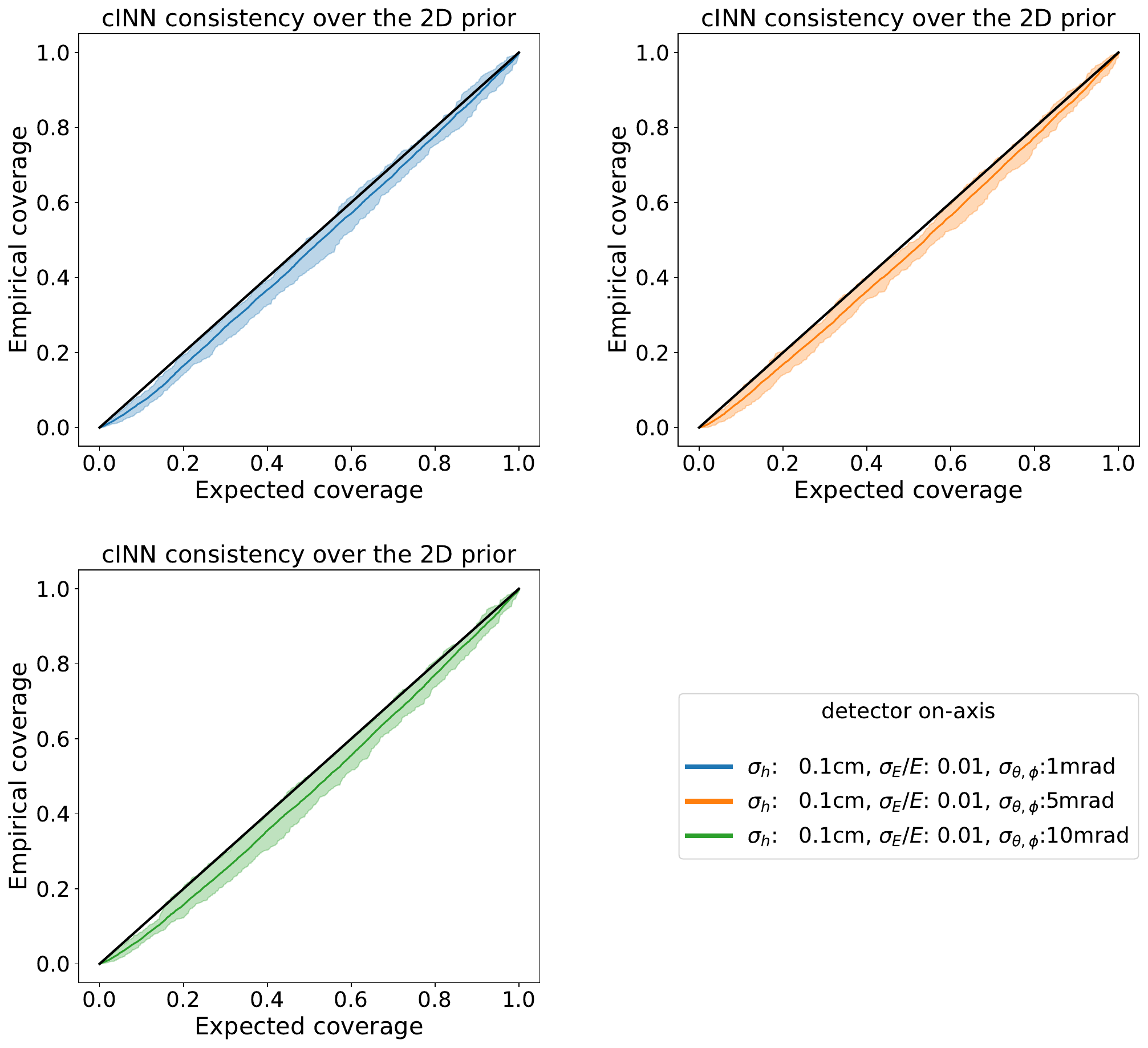}
    \caption{Distribution of the empirical coverage against the expected coverage. In black we can see the result in the case of correct posterior. 
    We have split the test samples in smaller sub-samples: the continuous colored line indicates the mean value over the sub-samples and the colored region encapsulates the empirical coverage values over all the sub-samples.}
    \label{fig:cons_2D}
\end{figure*}

In the limit where our posterior perfectly approximates the correct one and for large enough statistics, we would see that the empirical coverage and the expected coverage coincide. 
The results for our case are given in \cref{fig:cons_2D} for a representative selection of our networks. Our test samples consisting of 10000 sets have been split into 20 smaller sub-samples to evaluate the statistical uncertainty on the coverage. 
Given our cINN and the possibility of sampling directly from the 2D joint posterior, it is straightforward to evaluate the empirical coverage. 
We see a percent level underestimation of the coverage, consistent with the limited overfitting seen in the training of our networks. 

\begin{figure*}
    \centering
    \includegraphics[width=0.9\textwidth]{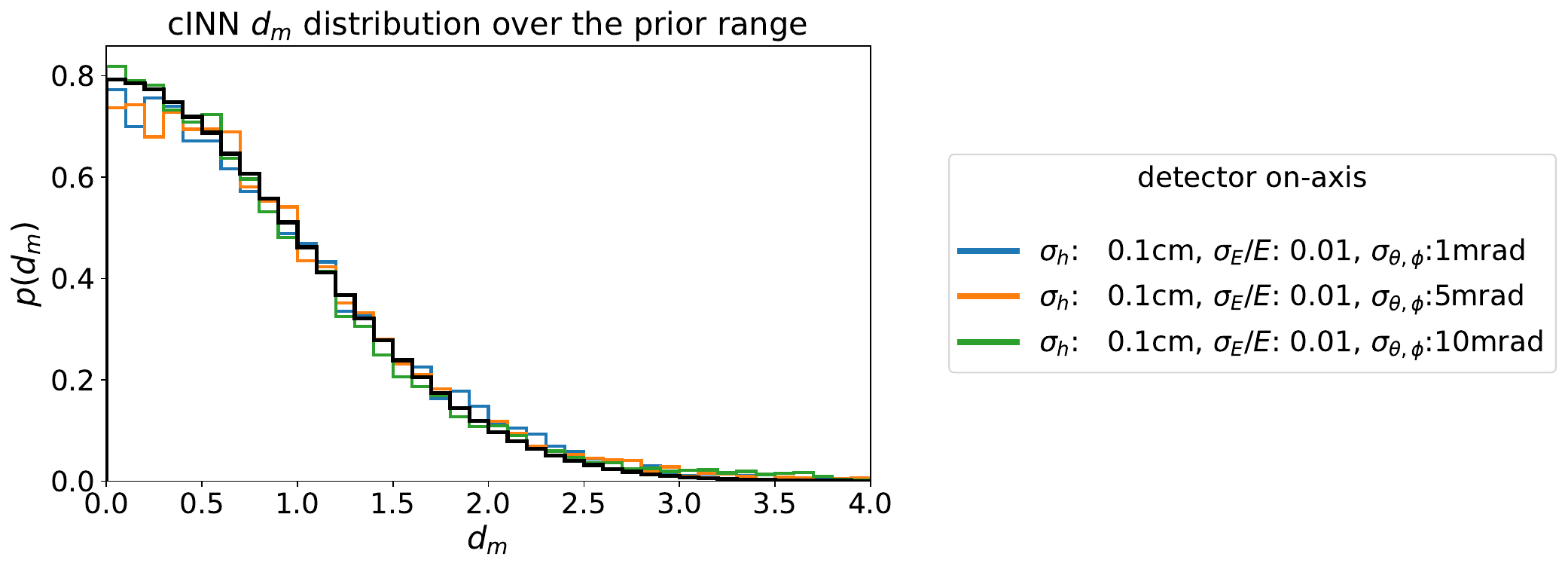}
    \caption{Distribution of the distance between estimated and true mass values in standard deviations. In black we show the comparison with the result from a Gaussian distribution. Even though the posterior is not required to be Gaussian, we find good qualitative agreement. We note, however, that larger differences may arise for worse detector resolution.}
    \label{fig:cons_dm}
\end{figure*}

The consistency has been evaluated over the whole prior and by considering the two-dimensional coverage. 
Not all regions of the parameter space are equally easy to constrain and in particular mass and (unboosted) lifetime present different challenges. 
Since in our scenario the performance is mainly given by the uncertainty on the mass, let us look at the distribution of $d_m=|\log_{10}\hat{m}-\log_{10}m_a|/\sigma_m$ in \cref{fig:cons_dm}. This is the distance between the true ALP mass value $m_a$ and our estimator $\hat{m}$ in units of standard deviations. 
With our (approximate) posterior we are able to derive for each test set both a mass estimator from the maximum of the posterior and a standard deviation value from the variance of the posterior samples. 
Even though our result is not expected to be Gaussian, it is instructive to show a comparison with the Gaussian result in black in \cref{fig:cons_dm}. 
We can see that the distribution given by the cINN closely resembles the normal distribution, although with slightly stronger tails.

\subsection{Dealing with background events}\label{sec:fake}

Experimental efforts are devoted to reducing  backgrounds as far as possible. 
However, we cannot guarantee in general that a sample of observed events is background-free. 
As discussed above, our cINN approach employs a summary network to synthesize the information from three events into two high-level observables. 
This procedure assumes that all three events are true signal events generated from the same underlying process, i.e.\ the decay of an ALP with fixed parameters $\boldsymbol{\theta}=(\log_{10}m_a, \log_{10}(c\tau_a/m_a))$. 
To ensure that this approach gives sensible results, one must check whether the three events are compatible with each other before combining them.

To achieve this goal, we need a different cINN, which is trained on individual events rather than sets of three events. The architecture (including the summary network) is the same as for sets of three events, with the only difference being the dimensionality of the input. We then obtain the estimated posterior $p(\boldsymbol{\theta}| \mathbf{x}^k)$ for each event $\mathbf{x}^k$ separately. From these results, the compatibility can be directly quantified. From each posterior we get an estimator of the model parameters
\begin{equation}
    \hat{\boldsymbol{\theta}}^k \equiv \argmax_{\boldsymbol{\theta}}\, \tilde p (\boldsymbol{\theta}| \mathbf{x}^k)\;,    
\end{equation}
and we evaluate their compatibility via:
\begin{equation}
C_{kl}\equiv 1- \int_{\tilde p(\boldsymbol{\theta}|\mathbf{x}^l)>\tilde p(\hat{\boldsymbol{\theta}}^k|\mathbf{x}^l)}\tilde p(\boldsymbol{\theta}|\mathbf{x}^l)\,\mathrm{d}\boldsymbol{\theta}.
\end{equation}
$C_{kl}$ is then the measure of compatibility between the posteriors. 
Explicitly, we consider the posterior $\tilde p(\boldsymbol{\theta}|\mathbf{x}^l)$ and evaluate the smallest credible region that contains $\hat{\boldsymbol{\theta}}^k$. 
Our compatibility measure is then one minus the corresponding credibility. 

\begin{figure*}
    \centering
    \includegraphics[width=\textwidth]{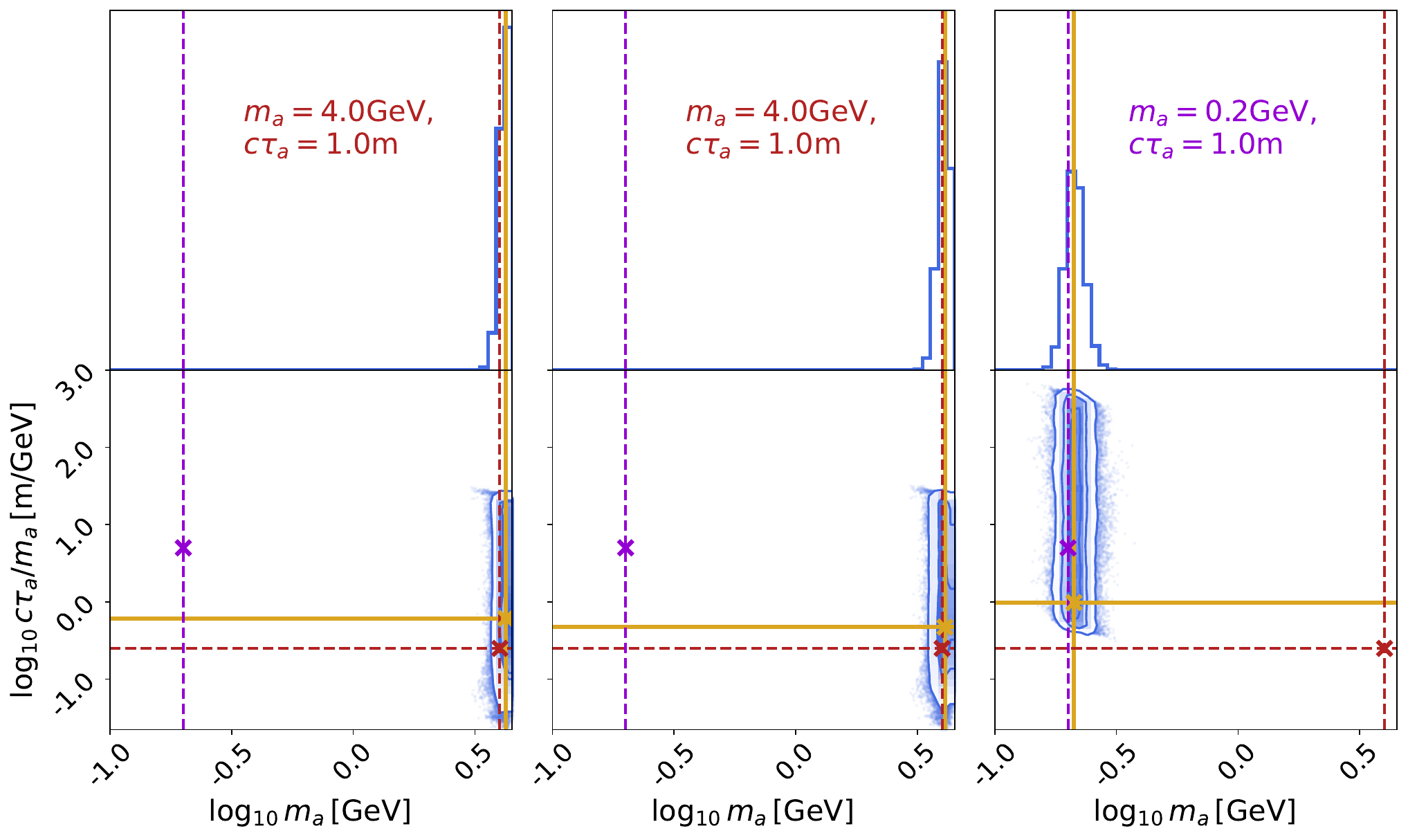}
    \caption{Estimated posterior from three events separately. The red cross indicates the parameter points generating the first two true events, the purple cross indicates the parameter points generating the last event (which we call background). The golden cross indicates for each of the posteriors the maximum posterior value $\hat{\boldsymbol{\theta}}^k$. The three contours indicate the $50\%$, $68\%$ and $95\%$ credible regions. The sampled posterior points outside of the $95\%$ credible region are also shown in the plot.}
    \label{fig:fake1_example}
\end{figure*}

To test this procedure, we generate samples that contain two true signal events (generated using the same ALP parameters) and one background event (generated using different values for the ALP mass and lifetime). 
The individual posterior for one such set of events is visualised in \cref{fig:fake1_example}. By considering single events, we see that our approach works also in this scenario. Increasing the number of seen events leads to narrower posterior, as expected. The main reason we focus on the scenario of multiple seen events is that in this case it is possible to assess their compatibility and so we have a procedure to identify background events. For instance here,
it is clear that the third event is not compatible with the first two by visual inspection. 
The compatibility measure $C_{kl}$ for the case shown in \cref{fig:fake1_example} is found to be
$$
C_{kl}=
\begin{pmatrix}
1 & 0.59 & 0 \\
0.97 & 1 & 0 \\
0 & 0 & 1
\end{pmatrix} \; .
$$
Here the vanishing off-diagonal elements in the third column and row clearly indicate the incompatibility of the third event with the first two.

\begin{figure*}
    \centering
    \includegraphics[width=0.66\textwidth]{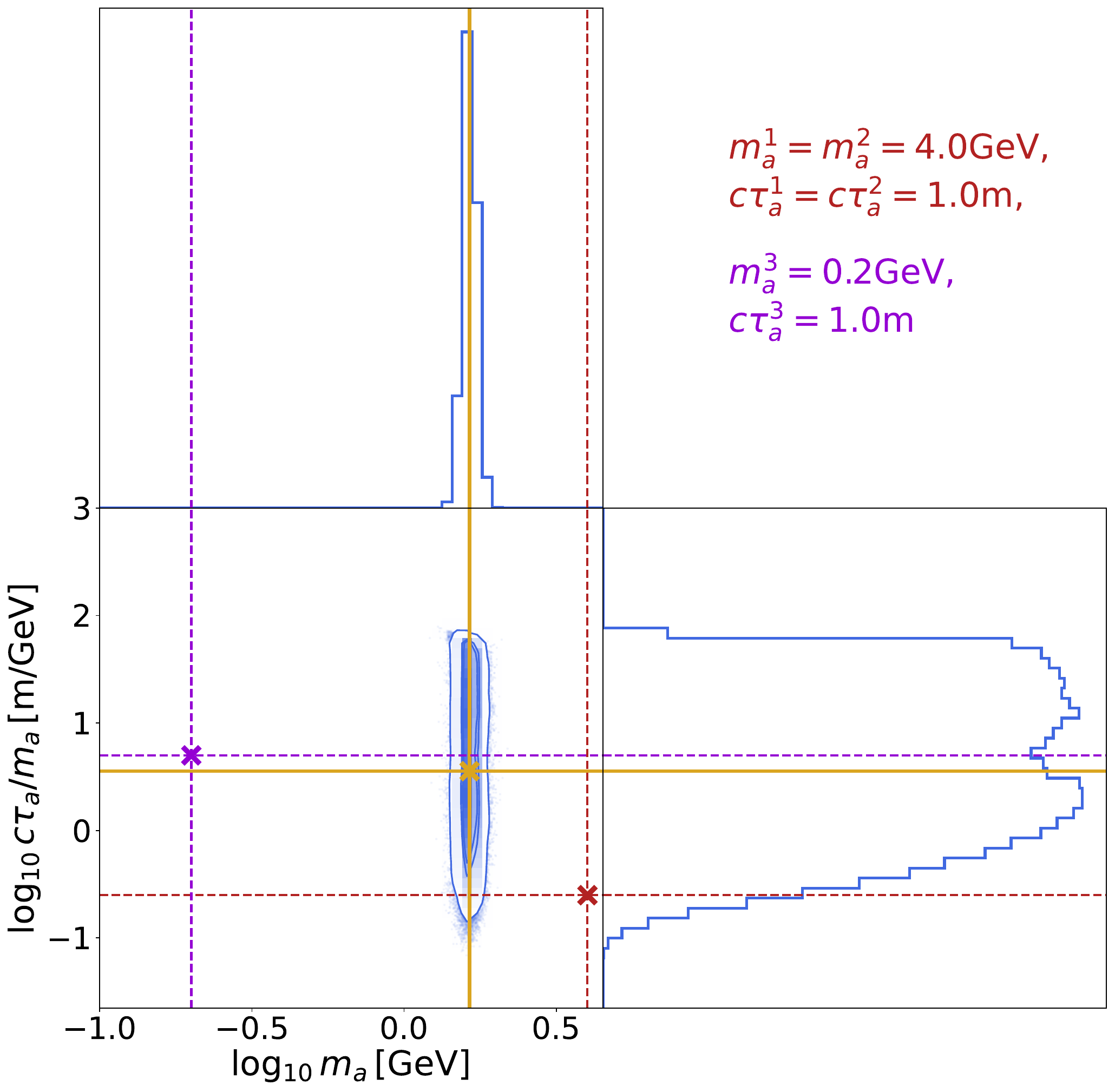}
    \caption{Estimated posterior from the three combined events. 
    The red cross indicates the parameter points generating the first two true events, the purple cross indicates the parameter points generating the last event (which we call background). 
    The golden cross indicates the maximum of the posterior.}
    \label{fig:fake3_example}
\end{figure*}

If we come to the conclusion that the three events are not compatible, we would not proceed and combine them as input for our full cINN. 
It is nevertheless interesting to see what happens if we do, and this is portrayed in \cref{fig:fake3_example}. 
Clearly, we would come to wrong conclusions about the mass.
Even worse, the network confidently claims a narrow posterior, which gives no indication that the events are incompatible. 
This is likely because the network has never seen incompatible events during its training, so it reconstructs the posterior as narrow as usual.\footnote{We have also found cases where the network would ignore one or two of the events and reconstruct the posterior from the remaining ones. However, it appears difficult to force the network to always focus on the compatible ones only and we did not study this further.}

\begin{figure}
    \centering
    \includegraphics[width=0.5\textwidth]{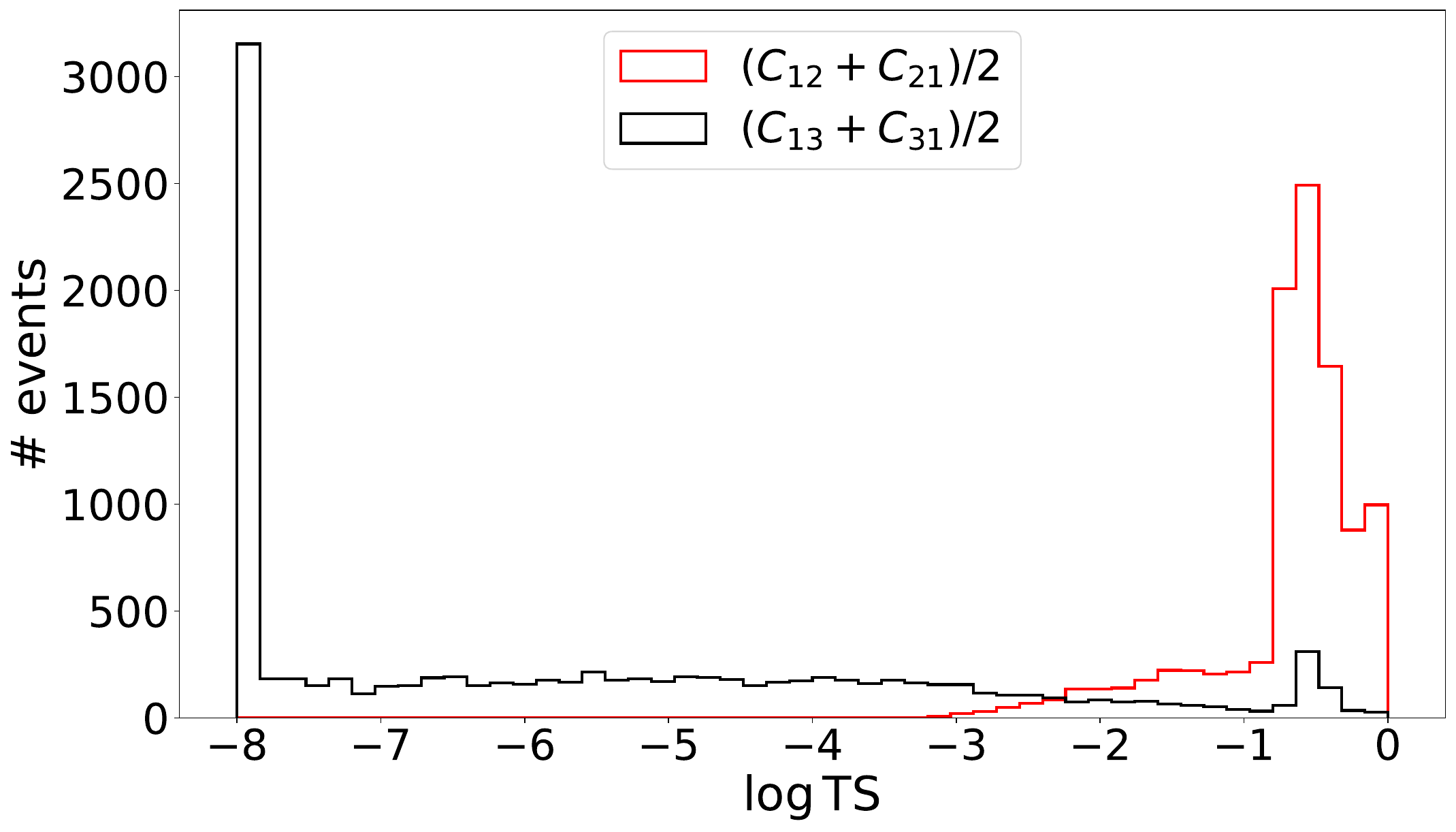}
    \caption{Distribution from pseudo-experiments of the compatibility test statistic for the case that the first two events are generated from the same parameter point and the third event is generated from a different point (see text for details). For a given signal acceptance (i.e.\ type I error rate), we can use this distribution to determine the background rejection (i.e.\ type II error rate).}
    \label{fig:Cij}
\end{figure}

Using the compatibility measure $C_{kl}$, it is straightforward to construct a test statistic (TS) to perform a hypothesis test of compatibility. 
Since the compatibility matrix is not symmetric, we consider the average $(C_{12} + C_{21})/2$ as TS for the compatibility of the first two events and $(C_{13} + C_{31})/2$ as TS for the compatibility of the first and the last event. 
To determine the distribution of each TS, we consider a sample containing sets of three events. 
For each set, event 1 and 2 have been generated from the model parameters $m_a=\SI{1}{\giga\electronvolt},\, c\tau_a=\SI{1}{\meter}$, while event 3 has been generated from $m_a=\SI{0.8}{\giga\electronvolt},\, c\tau_a=\SI{1}{\meter}$. 
We visualize the TS distributions in \cref{fig:Cij}. We observe that the TS behaves very differently depending on whether the two events are compatible or not, but we also find some overlap between the two distributions. 
Clearly, the degree of overlap depends on the parameter values used for the background event and on the detector setup. The more similar the events, and the poorer the detector resolution, the harder it will be to distinguish background and signal events.

In a realistic scenario, for each detector setup we would need to perform a Monte Carlo simulation to establish the signal efficiency at a fixed background rejection. A perfect background rejection is not possible as the overlap between background events and signal events is unavoidable for the background template we considered.  The question of establishing and maximizing the power of the experiment is compelling, but not easy to address. A classifier trained specifically to distinguish between compatible and incompatible events could outperform the simple approach outlined here in terms of signal acceptance. A detailed study of the experimental sensitivity in the presence of varying number of background and signal events is currently under study and will be presented in a future work. In this work we also address the question of establishing a criterium for event compatibility, depending on the desired background rejection and detector setup \cite{unpublished}.

Using the procedure outlined above, background events can be accurately identified and eliminated. 
If we come to the conclusion that all events in the set are compatible with each other, we would proceed to combine them. 
In principle this could be done by simply multiplying the individual likelihoods, or by combining the posteriors in an appropriate way. 
However, since we only have access to the approximate likelihoods/posteriors predicted by the cINN, doing so might amplify any inaccuracies. 
More accurate results are obtained by using the cINN trained on sets of three events, which is what we will do in the following.

To conclude this discussion, let us emphasize that there may be different ways to check compatibility of the events before combining them. 
In particular, it may be possible to construct a more powerful TS if the dominant source of background is known. 
The approach discussed above has the advantage that we can in principle use information from the full posterior to assess the probability of type~I and type~II errors, whereas for example a classifier would typically only yield a single number.

\subsection{Effect of the input features uncertainties}

So far we have focused on a fixed detector setup and discussed how the posterior looks like for different parameter values.
Let us now consider how the same event is seen by different detector setups, meaning different sets of uncertainties on the input variables. 
For each detector setup we train a corresponding cINN.
We use the same architecture and the same training hyperparameters for all of them: the underlying physical process is the same, so the algorithm adapts easily to different smearings of the input observables. 

Each detector setup is defined by the set of standard deviations $\sigma$ used for the Gaussian smearing that models the detector resolution. 
The same uncertainty is applied to $\{x_1, x_2, y_1, y_2\}$ as the resolution here is mainly given by the detector granularity.
We also assume that $\theta$ and $\phi$ have the same uncertainty, and that all uncertainties are uncorrelated.
As mentioned in \cref{sec:learningposterior}, we note that realistically energy and angular resolutions are function of the photon energy itself, with resolutions improving for increasing photon energies $\sigma(E)/E \approx 10-15\%/\sqrt{(E(\text{GeV})}$ and $\sigma(\theta)=\sigma(\phi) \approx 30-40\,\text{mrad}/\sqrt{(E(\text{GeV})}$.
For easier comparison between different detector setups, we use constant resolutions in our study, but we have checked that our methods work  identically for energy-dependent resolutions.
In a realistic detector with longitudinal segmentation, not all showers will start at the same $z$-position which is again conceptually straightforward to include in our cINN inputs.

The set of uncertainties is provided in \cref{tab:unc_setups} for a total of 9 detector setups and corresponding networks. For sufficiently good angular resolution, the calorimeter hit resolution plays no significant role and has therefore been fixed to $\sigma(h)=\SI{0.1}{\centi\meter}$. The results for the case of poor angular resolution and varying calorimeter hit resolution are discussed in \cref{sec:calo_hit}.

\begin{table}
    \centering
    \begin{tabular}{c|c}
        Feature uncertainty & Values scanned \\
        \hline
          $\sigma(E)/E$ & $[0.01, 0.05, 0.1]$ \\  
          $\sigma(h)$ & $[0.1\SI{}{\centi\meter}]$ \\
          $\sigma(\theta),\sigma(\phi)$ & $[\SI{1}{\milli\radian}, \SI{5}{\milli\radian}, \SI{10}{\milli\radian}]$ 
    \end{tabular}
    \caption{Summary of the detector setups considered. 
    Each uncertainty value is combined with each of the other uncertainty values for a total of 9 combinations.
    A complementary case for large angular uncertainties and varying calorimeter hit resolution can be found in \cref{sec:calo_hit}.}
    \label{tab:unc_setups}
\end{table}

\begin{figure*}
    \centering
    \includegraphics[width=0.8\textwidth]{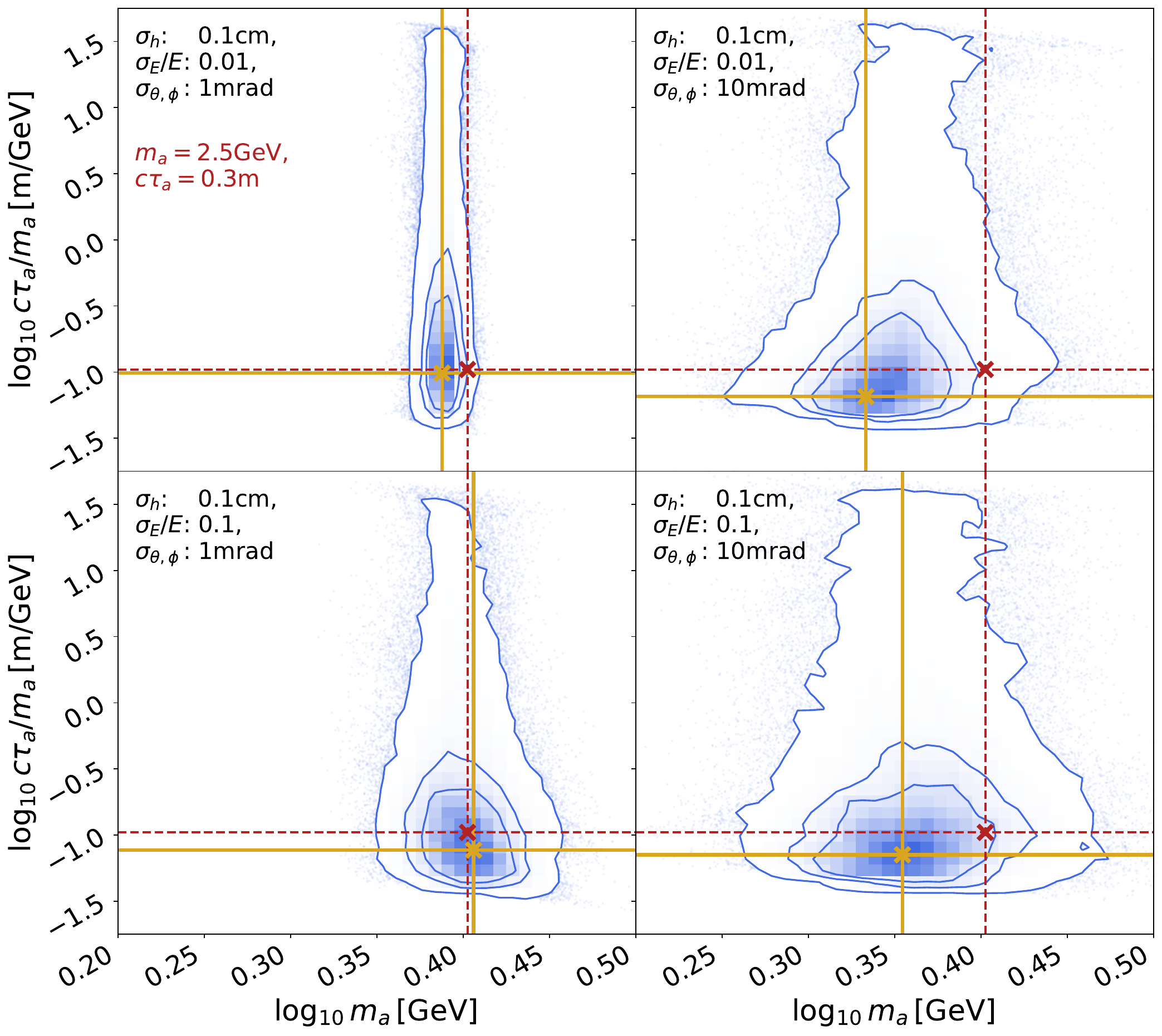}
    \caption{Comparison between the joint posteriors on the same set of events, but for different detector setups. In this figure we keep the calorimeter hit resolution fixed, while we vary the energy resolution and the angular resolution. The figure does not show the whole prior range, but is zoomed on the region where the posterior is non-zero. The three contours indicate the $50\%$, $68\%$ and $95\%$ credible regions. The sampled posterior points outside of the $95\%$ credible region are also shown in the plot.}
    \label{fig:comparison}
\end{figure*}

To each of these detector setups corresponds an estimated posterior $\tilde p(m_a, c\tau_a|\mathbf{x})$ which differ in the uncertainties assigned to $\mathbf{x}$.  To first approximation, as we increase the input features uncertainties, we expect the posterior to get broader. The inferred posteriors can hence be used to draw conclusions about the performance of the detector setup. Before comparing the different detector setups in a quantitative way in the next section, let us briefly visualize how the posterior on the model parameters broadens as we increase the uncertainty on the low level observables in \cref{fig:comparison} for a fixed set of events.

As expected, the spread of the posterior increases as the detector resolution on the energy and/or the angles deteriorates. As discussed above, it is hard to pin down the ALP lifetime even in the best-case scenario, and the spread of the marginal posterior in the lifetime does not change significantly from the best to the worst case scenario. The reason for this can be understood when looking at the ideal scenario where the vertex is perfectly reconstructed. The decay vertex position is mainly determined by the particle lifetime, but also by the requirement that both produced photons hit the calorimeter with sufficient separation. We can use our training dataset to fit the joint distribution of vertex position and unboosted lifetime. However, even in the case the exact vertex position is known, it is not possible to extract significant information on the particle lifetime. This results in roughly flat likelihoods and posteriors closely following the prior. Thus, regardless of the detector setup or inference procedure, we will recover a roughly constant uncertainty on the unboosted lifetime. As mentioned earlier, the situation would change if we  considered an unrealistically large decay volume, which in general leads to more peaked posteriors in the lifetime.

The width of the marginal posterior on the mass, on the other hand, does exhibit significant changes with the detector resolution. But even in the case of poor energy and angular resolution, we can estimate the ALP mass with relative low uncertainty. Indeed, in this case the cINN significantly outperforms the diphoton invariant mass, as we will see in the next section. 

\section{Results}\label{sec:results}

\subsection{Performance evaluation: on-axis}\label{sec:performance_on}

We are now ready to discuss how we evaluate the detector performance for a given setup. In our scenario, three signal events have been observed by the experiment and we want to use these three events to infer the model parameters. The best detector for this purpose will be the one that has the highest model discrimination. Qualitatively speaking, the more non-overlapping posterior surfaces we can fit into the parameter space, the better we are at discriminating models. In other words, the posterior should be as narrow as possible, so the area enclosed by it should be small. 

To evaluate the performances, we build three test datasets. Each of these datasets contains 10000 samples, and they are all generated for a fixed lifetime of $c\tau_a=\SI{1}{\meter}$, 
and three different ALP masses: a low ALP mass of $\SI{200}{\mega\electronvolt}$,  a medium ALP mass of $\SI{1}{\giga\electronvolt}$ and a large ALP mass of $\SI{4}{\giga\electronvolt}$.

In the following, we compare the cINN performance with the standard approach of using the diphoton invariant mass 
\begin{equation}
    m_{\gamma\gamma}^2\equiv (p_{\gamma_1}^\mu+p_{\gamma_2}^\mu)^2=(E_1+E_2)^2 -  |\mathbf{p}_1+\mathbf{p}_2|^2
\end{equation}
where $p_{\gamma_i}^\mu$ and $\mathbf{p}_i$ are respectively the photon 4-momenta and 3-momenta. 
The diphoton invariant mass can be interpreted as a simple and powerful analytic mass regressor. Since our observation consist of three events, we will take as mass estimator the average of the three diphoton invariant masses.
Like for any regressor, this procedure will only return an estimate for the ALP mass and not its uncertainty.
In order to estimate the mass uncertainty for a single set of three events, we apply the smearing several times. 
In this way we will get a collection of several imperfect events coming from the same truth values. 
Once we have this collection of events, we can apply the regressor to obtain a distribution of masses. 
The standard deviation of this distribution then measures the uncertainty on the inferred mass. 
The same procedure could be applied when using a NN regressor trained to reconstruct the mass.

We note that, while the diphoton invariant mass is expected to yield an optimal mass estimator for detectors with very high energy and angular resolution, for other detector setups it may be possible to improve the regressor further, for instance by applying vertex fitting to correct the photon directions. 
We have found no qualitative differences when using improved analytic regressors or NN regressors trained on the same datasets used by the cINN. 
A detailed study of alternative possibilities is outside the scope of the current work.

\begin{figure*}
    \centering
    \includegraphics[width=0.9\textwidth]{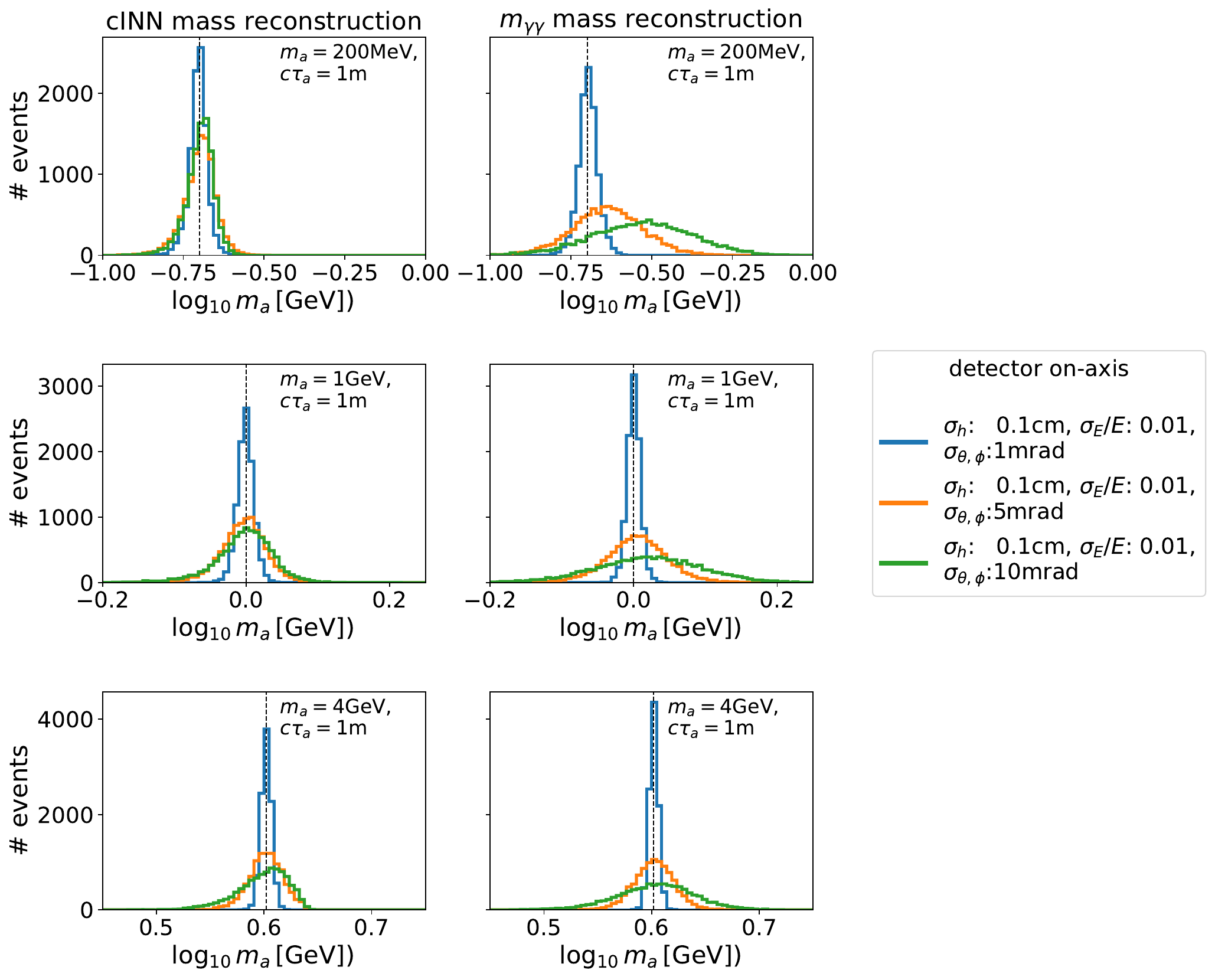}
    \caption{Mass reconstruction performed by the cINN (left) and by using the diphoton invariant mass (right). Different colors correspond to different detector setups (as indicated in the legend), while the three rows correspond to the ALP masses $m_a = 200 \, \mathrm{MeV}$, $1\,\mathrm{GeV}$ and $4\,\mathrm{GeV}$. 
    Vertical black lines indicate the true mass values. 
    Different bin widths have been used for the three benchmark masses.}
    \label{fig:mass_spread}
\end{figure*}

Let us now evaluate the cINN and the diphoton invariant mass on our test datasets and look at the distribution of the mass estimators in \cref{fig:mass_spread}. 
First we consider three detector setups of varying angular resolution, which turns out to be the most important detector property. 
The three detector setups are represented by lines of different colour, while the three rows correspond to the three different values of the ALP mass. 
These plots are indicative of the performance, but most importantly they are useful to identify possible biases. 
We can see that the cINN is generally unbiased for all our benchmarks, while for detectors with poor angular resolution the diphoton invariant mass exhibits a bias towards larger masses especially for small ALP masses. 

In principle, one could use the distribution of estimated masses over the entire test dataset to extract the variance in the mass estimator. 
However, to properly evaluate the performance, it is preferable to evaluate $\sigma_m=\sigma (\log_{10}\hat{m})$ for each sample separately, because this makes it possible to also evaluate the variance of $\sigma_m$ over the dataset. 
We also point out that the error on the logarithm of the ALP mass is related to the relative error on the ALP mass, such that we can directly compare our results for different assumptions on the true ALP mass. 
Given the number of different detector setups that we consider, it is difficult to visualize the distributions of $\sigma_m$ for all cases. 
In the following we will therefore focus on $m_a=\SI{1}{\giga\electronvolt}$ while taking some representative detector setups which highlight our conclusions. 
The distributions for the other masses are provided in \cref{sec:appendix_plots}. 

\begin{figure*}
    \centering
    \includegraphics[width=0.9\textwidth]{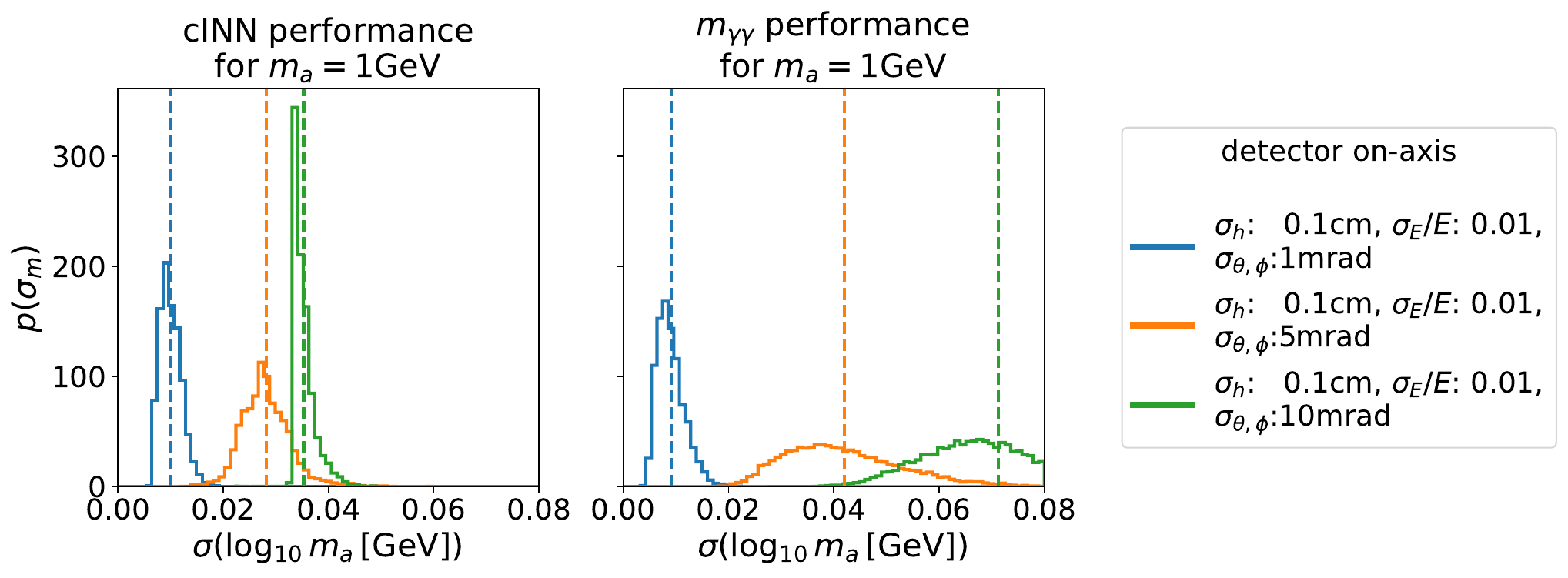}
    \caption{Mass uncertainty distribution over the 10k samples of the test dataset varying the angular resolution of the detector for $m_a=\SI{1}{\giga\electronvolt}$.}
    \label{fig:perf_angle_on}
\end{figure*}

In \cref{fig:perf_angle_on} we compare the performances for different detector setups, focusing again on the impact of changing the angular resolution. 
By showing the distribution of $\sigma_m$, this figure adds new information with respect to \cref{fig:mass_spread}, which showed the distribution of $\hat{m}$. 
As expected, we see that the ML approach does not perform better than the diphoton invariant mass in the case of good angular resolution. 
The distributions of $\sigma_m$ from cINN and diphoton are in this case very similar, indicating that the network has learnt to reconstruct exactly this high-level observable. This finding also enables us to assess the accuracy of the posteriors obtained from the cINN. For a detector with good angular resolution, we expect the marginal posterior on the ALP mass to follow the same distribution as the diphoton invariant mass. The fact that the $\hat{m}_a$ distribution given by the cINN agrees with the diphoton invariant mass distribution therefore indicates that the cINN performs well and yields accurate posteriors.
The situation changes when the resolution on the angles is decreased, as in this case the cINN can do significantly better than the naive $m_{\gamma\gamma}$.

\begin{figure*}
    \centering
    \includegraphics[width=0.9\textwidth]{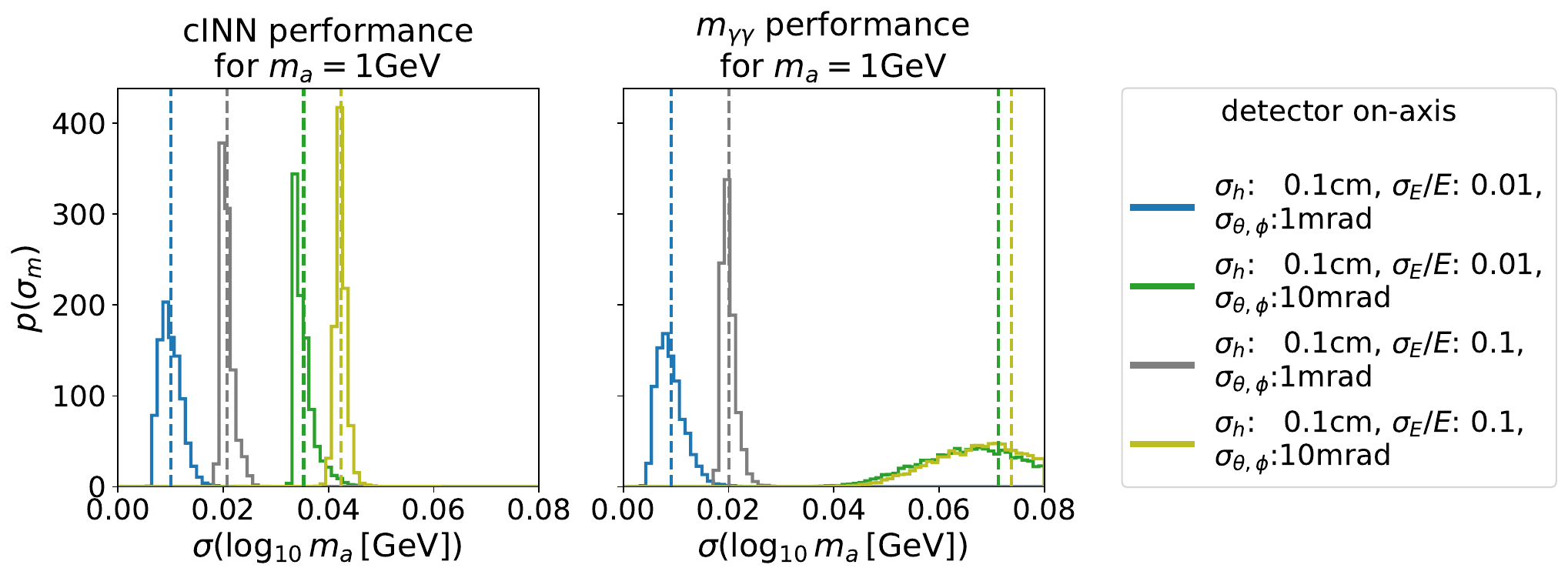}
    \caption{Mass uncertainty distribution over the 10k samples of the test dataset varying energy and angular resolution for $m_a=\SI{1}{\giga\electronvolt}$.}
    \label{fig:perf_angen_on}
\end{figure*}

Our analysis shows that the angular resolution of the detector is a major factor for our ability to constrain the ALP mass. The effect of also changing the energy resolution is investigated in \cref{fig:perf_angen_on}. 
For both values of the energy resolution considered, we find that, as long as the angular resolution is sufficiently good, the distributions of $\sigma_m$ for the cINN and for the diphoton invariant mass are very similar.  
This finding suggests that also for poorer energy resolution, the diphoton invariant mass remains the most informative high-level observable. 
However, the conclusion changes when we decrease the angular resolution. 
In this case the cINN performs better for both values of the energy resolution. 
Moreover, the cINN distribution shifts considerably when changing the energy resolution, while the $m_{\gamma\gamma}$ distribution is basically unaffected. 
This observation suggests that the cINN uses additional information contained in the photon energies, and consequently an improvement in the energy resolution helps the inference process.

\subsection{Performance evaluation: off-axis}\label{sec:performance_off}

In our discussion so far we have neglected the role of backgrounds in the sense that we have assumed that the experimental collaboration is able to provide a background-free sample. 
If that is not the case we have designed a test statistic that can be extracted from our network in order to diagnose the possible presence of background events. 
In reality, substantial experimental efforts are devoted to reducing the number of background events as much as possible.
One of the possible ways in which this can be achieved is by placing the decay volume and calorimeter off-axis with respect to the proton beam line. 
This possibility is interesting to investigate with our approach, since the different geometry implies different particle kinematics.
Here we will take inspiration from the SHADOWS proposal~\cite{Alviggi:2839484} and consider a displacement of $x_\text{cal}=\SI{2.25}{\meter}$, such that the edge of the decay volume is at a distance of $\SI{1}{\meter}$.

Displacing the detector affects not only the backgrounds but also the distribution of signal events. 
ALPs produced off-axis typically have smaller energies than those produced on-axis. 
This also leads to larger separation between the calorimeter hits. 
In combination these effects imply that an off-axis detector is sensitive to somewhat different regions of parameter space (i.e.\ longer ALP lifetimes) and generally exhibits a better performance in terms of the ALP mass reconstruction. 
Rather than comparing the two different detector positions in terms of the reconstruction performance, we will instead repeat the analysis from above, and explore the performance of an off-axis detector for varying detector resolutions.

Given the similarity of this geometry with the previous one, we do not need to adapt the network architecture or the hyperparameters. 
However, we will need to generate new training samples and re-train our network. 
We can then apply our cINN to new test datasets generated for the off-axis geometry. 
We consider the same benchmark points as before, but emphasize that the off-axis geometry inherently has a different sensitivity to them compared to the on-axis geometry. 

\begin{figure*}
    \centering
    \includegraphics[width=0.9\textwidth]{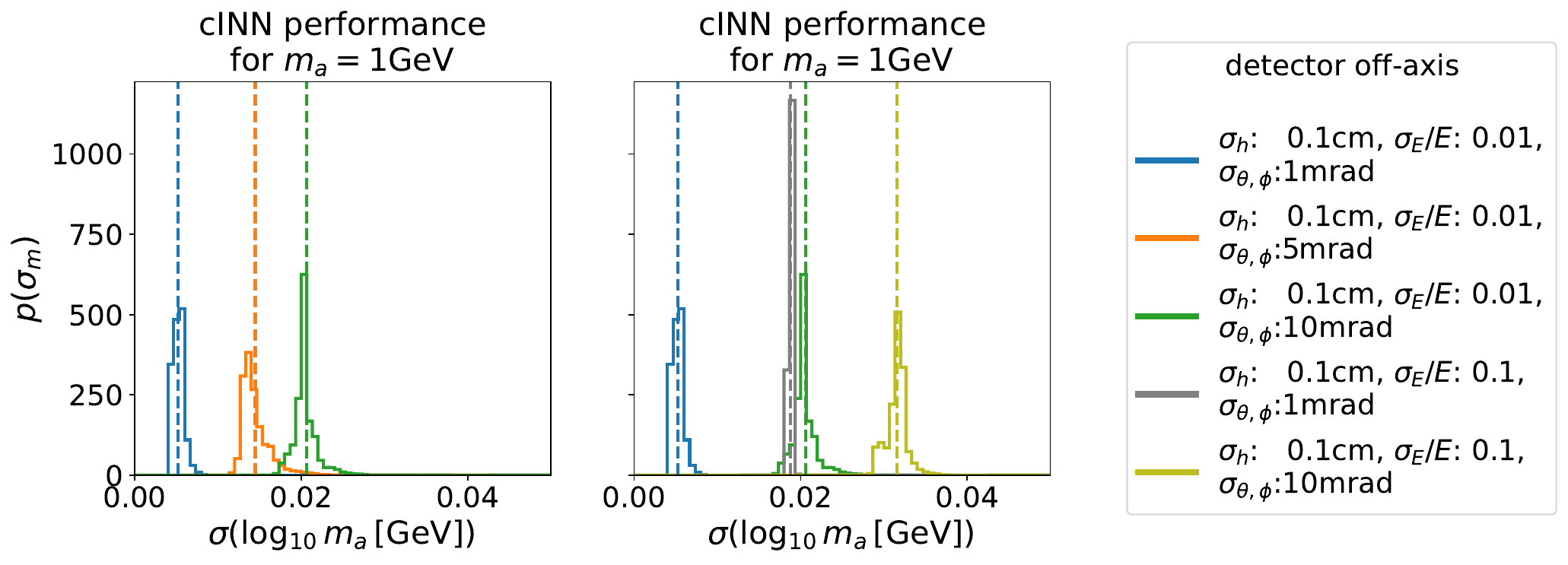}
    \caption{Uncertainty distribution over the 10k samples of the test dataset varying the only angular resolution (left) or both the angular resolution and energy (right) for the case of an off-axis detector for $m_a=\SI{1}{\giga\electronvolt}$.}
    \label{fig:perf_angle_off}
\end{figure*}

As we did in \cref{fig:perf_angle_on,fig:perf_angen_on} for the case of an on-axis detector, we can visualize the performance at varying angular and energy resolution for an off-axis detector in \cref{fig:perf_angle_off}. 
We drop the comparison with the diphoton invariant mass, as the conclusion is the same as before: for the best angular resolution our cINN reproduces the analytic regressor result, while it outperforms the latter for poor angular resolution.

In the left panel of \cref{fig:perf_angle_off}, we keep high energy resolution and vary the angular resolution, which is found to still play a major role. 
However, even in the case of $\sigma(\theta)=\sigma(\phi)=\SI{10}{\milli\radian}$ we can reconstruct the ALP mass with low relative uncertainty. When considering variations in both the angle and energy resolution in the right panel of  \cref{fig:perf_angle_off}, we observed a similar behaviour as for the on-axis case, but with an even stronger effect. In the on-axis case we saw that given an angular resolution of $\SI{10}{\milli\radian}$ decreasing the energy resolution increased the uncertainty on the logarithm of the mass by $25\%$ (dark green and light green curves in \cref{fig:perf_angen_on}). The same variation in the off-axis case leads to an increase in the logarithmic mass uncertainty by $50\%$. We conclude that different aspects of the detector resolution are important for different geometries. It is therefore essential to understand for each possible geometry individually which variables are most useful to infer the model parameters. The distributions obtained from the cINN analysis are an important diagnostic tools for this goal.

\subsection{Performance comparison}\label{sec:perf_comp}

To conclude our discussion, let us summarize the performance comparison for the different detector setups in a compact way. 
As before, we consider the three benchmark masses $m_a=\SI{0.2}{\giga\electronvolt},$ $\SI{1}{\giga\electronvolt},$ $\SI{4}{\giga\electronvolt}$, while for the lifetime we will always keep $c\tau_a=\SI{1}{\meter}$. 
We consider separately the on-axis and off-axis geometries and compare the 9 detector setups summarized in \cref{tab:unc_setups} with the calorimeter hit resolution fixed to $\SI{0.1}{\centi\meter}$. 
Results for the case of worse angular resolution ($\sigma(\theta) = \sigma(\phi) = \SI{100}{\milli\radian}$) and worse resolution of the calorimeter hit positions can be found in \cref{sec:calo_hit}. 

\begin{figure*}
    \centering
    \includegraphics[width=\textwidth]{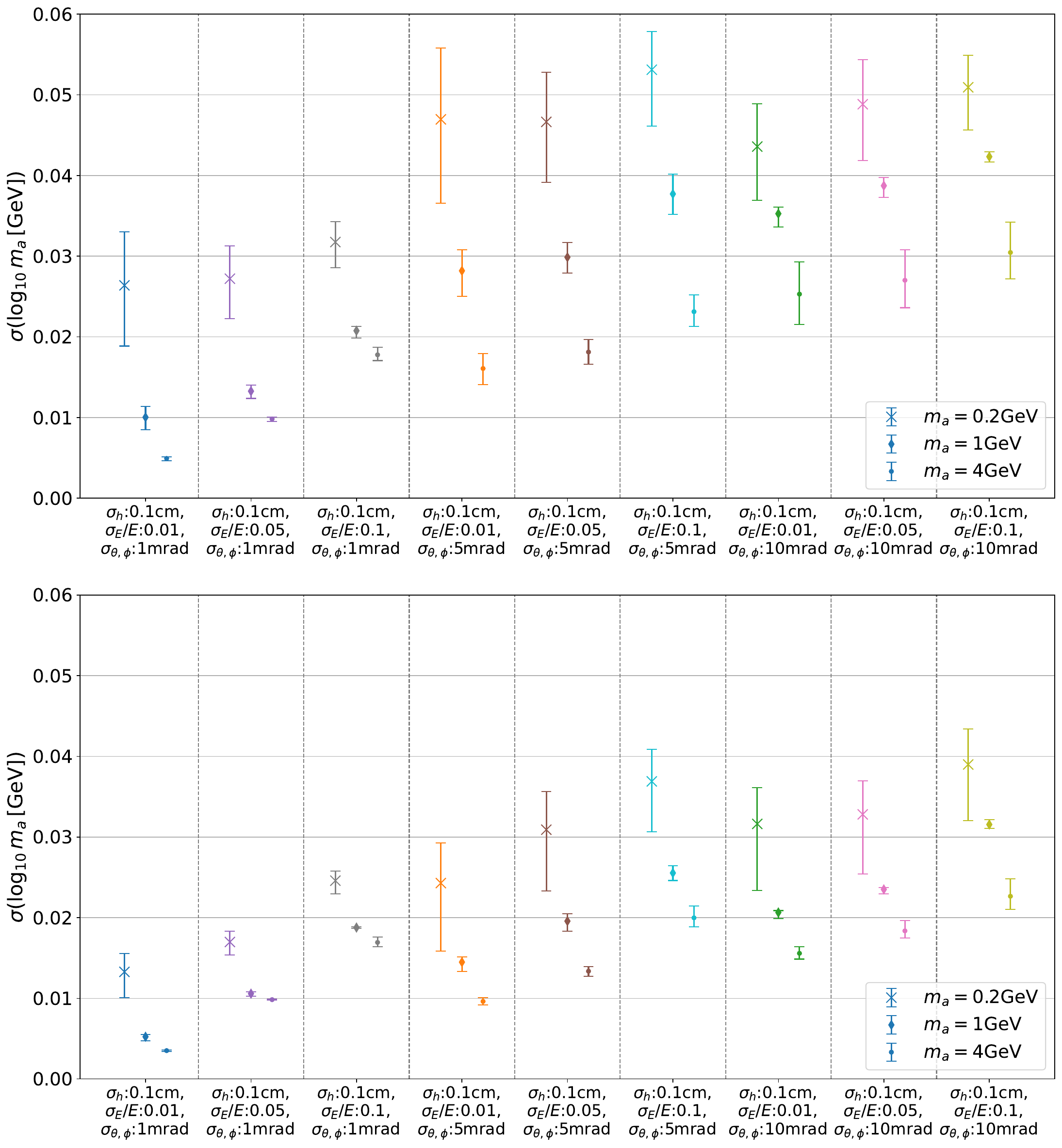}
    \caption{Compact performance comparison for fixed calorimeter hit resolution for on-axis (upper plot) and off-axis geometry (lower plot). Different colors correspond to different detector setups and different markers are used for the benchmark masses. The vertical bands indicate the 25 and 75 percentile of the distribution.}
    \label{fig:compact_perf_comp_displ}
\end{figure*}

The comparison of the different detector setups is shown in \cref{fig:compact_perf_comp_displ} for the case of an on-axis detector~(upper plot) and of an off-axis detector~(lower plot). 
We observed two general trends: Larger ALP masses are easier to constrain, and (apart from a few exceptions) detectors that perform better at constraining larger masses also perform better at constraining lower masses. 
The exceptions to this rule seem to indicate that constraining lower masses favors angular resolution over energy resolution.
We furthermore conclude that the angular resolution plays a major role in being able to constrain the  ALP mass, but its relevance also depends on the available energy resolution and the specific mass under consideration. 
For instance we can observe that having a good angular resolution is most relevant when we also have a good energy resolution. 
While the same trends are present for both on-axis and the off-axis detectors, there are some quantitative differences in the relation between detector resolution and reconstruction performance, suggesting that the relative importance of angular and energy resolution may be different for the two cases.

Comparison plots like \cref{fig:compact_perf_comp_displ} not only allow to compare different detector setups (where unsurprisingly the detector with best resolution is the best at constraining the ALP mass), but also to quantify by how much. 
This means that it is possible to use this type of plot to understand whether it is worth or not to invest more resources toward improving the resolution for measuring a specific kinematic variable. 
Conversely, if we want to measure the ALP mass with a given relative uncertainty, we can understand which detector setups would achieve that. 

At first sight, a naive comparison of the top and bottom panel of \cref{fig:compact_perf_comp_displ} suggests that moving the detector off-axis improves the detector performance. 
This finding likely reflects the fact that ALPs produced at an angle relative to the beam direction typically have smaller boost factors, which leads to larger photon opening angles and facilitates the reconstruction of the underlying process. 
At the same time, the distribution of transverse momenta carries information about the ALP mass, which may be more easily extracted from an off-axis experiment. 
We emphasize however that moving the detector off-axis leads to a significantly smaller signal acceptance, such that one should really compare the performance for a different number of observed events in the two cases. 
We expect that $\sigma_m$ scales approximately proportional to $1/\sqrt{n}$ with the number $n$ of observed events, such that the different performances shown in \cref{fig:compact_perf_comp_displ} could be easily compensated by increasing $n$ by a factor of 2--4.

Finding the best detector placement and resolution then becomes a difficult optimisation problem, which also needs to include the expected number of background events and their distribution. 
In practice, one would also vary additional parameters, such as the distance and length of the decay volume and the transverse size of the detector. 
Since considering all of these possibilities is beyond the scope of this work, the two panels of \cref{fig:compact_perf_comp_displ} cannot be directly compared in a meaningful way.

\section{Conclusions}

The \textit{inverse problem} refers to the challenges and limitations of performing parameter inference from experimental data for physical scenarios of interest. Usually, this problem is considered in the context of constructing optimal high-level observables for existing experiments. In the present work we have instead considered the inverse problem in the context of experimental design, i.e.\ we have compared different detector setups in terms of their performance with respect to parameter inference. Doing so requires a fast and adaptable way to perform inference while varying the experimental properties.

Our scenario of interest is the detection of ALPs decaying to photons in proton beam-dump experiments. 
This scenario is not only well-motivated from the physical and experimental point of view, but it also illustrates the key challenges of parameter inference. 
For beam dumps with very large decay volumes, photon energy and direction resolution are generally not good enough to directly infer the decay vertex and the invariant mass of the decaying particle with high precision. 
To infer the underlying ALP parameters, one then needs to include  additional information from other kinematic variables. In such a case, the interplay between the different observables will depend on the specific angular and energy resolution of the detector under consideration.

In this work we have demonstrated that in spite of these difficulties, conditional invertible neural networks are able to accurately reconstruct the posterior of the ALP model parameters for a given detector setup without the need for complex network architectures or explicit physical intuition~\cite{Andreassen:2018apy,Lai:2020byl,Bogatskiy:2023nnw}. 
For detectors with limited resolution, these networks significantly outperform conventional approaches, such as reconstructing the ALP mass from the invariant mass of the photon pair, suggesting that the conditional invertible neural network learns to extract additional information from the ALP distribution. 
The speed and adaptability of this machine-learning algorithm allow for the comparison of different detector setups, thus addressing the inverse problem already at the stage of experimental design. In our work we have focused our attention on a simplified simulation. However, the framework we have discussed directly extends to more accurate simulations. Not only that, but other sources of uncertainty not considered in this work can be integrated in the framework by suitably modifying the input features.

To obtain robust results, it is essential that we can trust our neural networks to perform correct inference and that we can trust the experimental observation to not be contaminated by background events. 
We address the first issue by comparing the empirical coverage against the expected coverage (see \cref{fig:cons_2D}). This comparison demonstrates that on average the cINN correctly determines the model parameters and their uncertainties. Moreover, we show that in the case of good angular and energy resolution the mass estimate and uncertainty obtained from the cINN agree with the ones obtained from the diphoton invariant mass distribution.
To address the second issue, we consider sets of three signal events, which enables us to check the compatibility between them. 
We have proposed a test statistic derived from the same cINN that evaluates the posterior to confirm that there are no background events in a given set of events.

In order to evaluate the inference power and hence the performance of a given detector, we consider the width of the posterior on the model parameters, i.e.\ the ALP mass and lifetime. 
We have shown that (at least for experiments with a relatively short decay volume), it is possible to avoid degeneracies between the two parameters, which makes it possible to integrate over the ALP lifetime and focus on the marginal posterior for the mass.
The width of this posterior, $\sigma_m=\sigma (\log_{10}\hat{m})$, then serves as performance measure for the different detector setups. 

In our analysis we have considered a total of 18 different detector resolutions (varying independently the energy resolution, angular resolution and position resolution) as well as two detector geometries that differ in their displacement $x_\text{cal}$ relative to the beam axis. 
The performances of the different detector setups are summarized and visualized in \cref{fig:compact_perf_comp_displ}. 
This figure illustrates the way in which our approach can be used to guide experimental design and explore the interplay and trade-offs between different aspects of detector resolution and geometry. 
We emphasize, however, that these plots do not include the effect of changing the detector geometry on the signal acceptance and the background suppression and therefore should not be directly compared to each other. 
Moreover, the detector geometries and experimental uncertainties have been simplified and do not necessarily represent the performance of realistic experiments. 
Nevertheless, it is possible to quantify the role of the detector resolution for different geometries with a single algorithm, highlighting the adaptability of our approach. 

The idea to search for feebly-interacting particles using new beam-dump experiments with state-of-the-art detectors is rapidly gaining momentum in the community. 
As different proposals are being put forward that vary in many aspects from the beam type over the detector geometry to the specific instrumentation, it becomes essential to understand which avenue promises the greatest gain of knowledge from a successful detection. 
The goal of the present work is to provide a consistent, fast and adaptable algorithm to facilitate this discussion and allow for the comparison of experimental proposals. 
The next step will be to apply this framework to realistic proposals and background distributions in order to optimize detector design and guide the experimental program.

\section*{Acknowledgements}

It is a great pleasure to thank (in alphabetical order) Alexander Heidelbach, Jan Kieseler, Markus Klute, Gilles Louppe and Kylian Schmidt for discussions, Giovani Dalla Valle Garcia for collaboration at early stages of this project, and Kierthika Chathirathas for feedback and comments on earlier versions of the manuscript. 
AM would like to thank Christoph Weniger and all the participants of the Swyft workshop for the very insightful discussions about simulation-based inference.
This work was supported by the Helmholtz-Gemeinschaft Deutscher Forschungszentren (HGF) through
the Young Investigators Group VH-NG-1303, and by
the Deutsche Forschungsgemeinschaft (DFG) through
the Emmy Noether Grant No. KA 4662/1-2 and grant
396021762~--~TRR~257. 

\section*{Code and data}

The code used for the event generation, the network architecture and its training can be found at \url{https://github.com/amorandini/SBI_axion}. An example Jupyter notebook which explains how to derive the plots and results in this paper is also provided. 
Neural networks are built and trained with Tensorflow \cite{tensorflow2015-whitepaper}, Tensorflow~Probability \cite{DBLP:journals/corr/abs-1711-10604} and use Keras \cite{chollet2015keras} as backend. 
The 2D posterior plots make use of corner.py \cite{corner}.

\appendix

\begin{figure*} [b]
    \centering
    \includegraphics[width=0.8\textwidth]{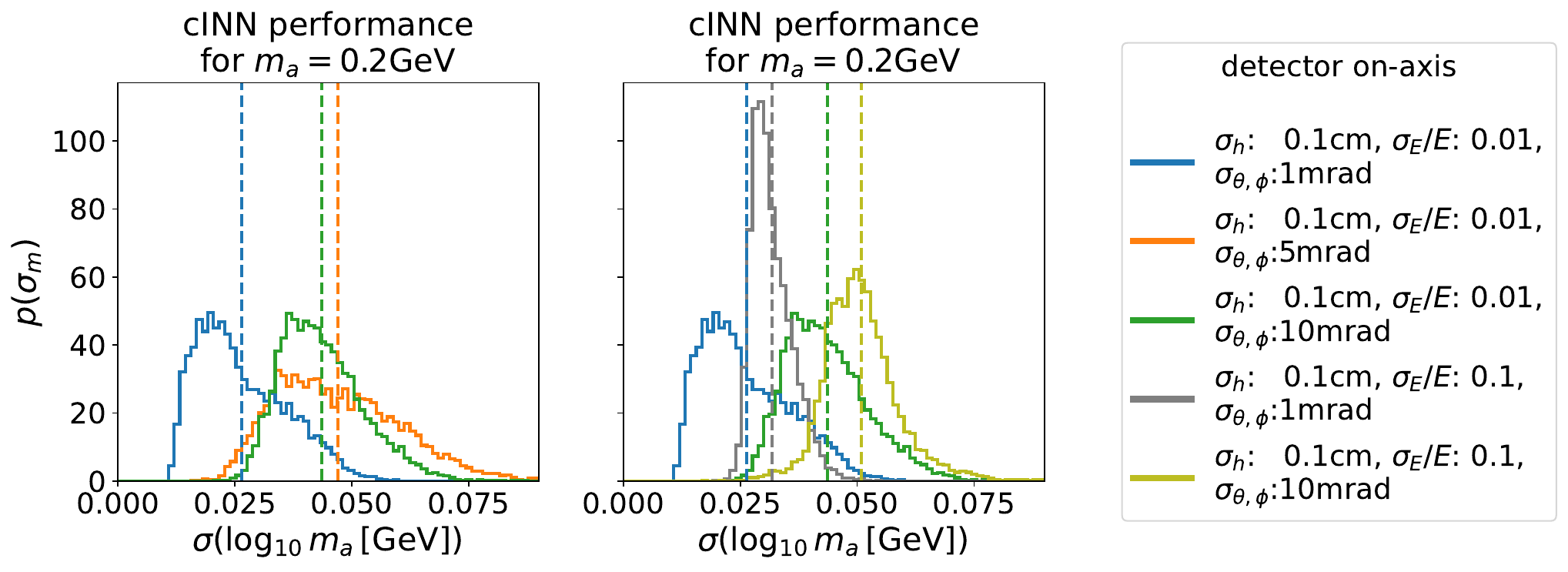}
    
    \vspace{2mm}
    \includegraphics[width=0.8\textwidth]{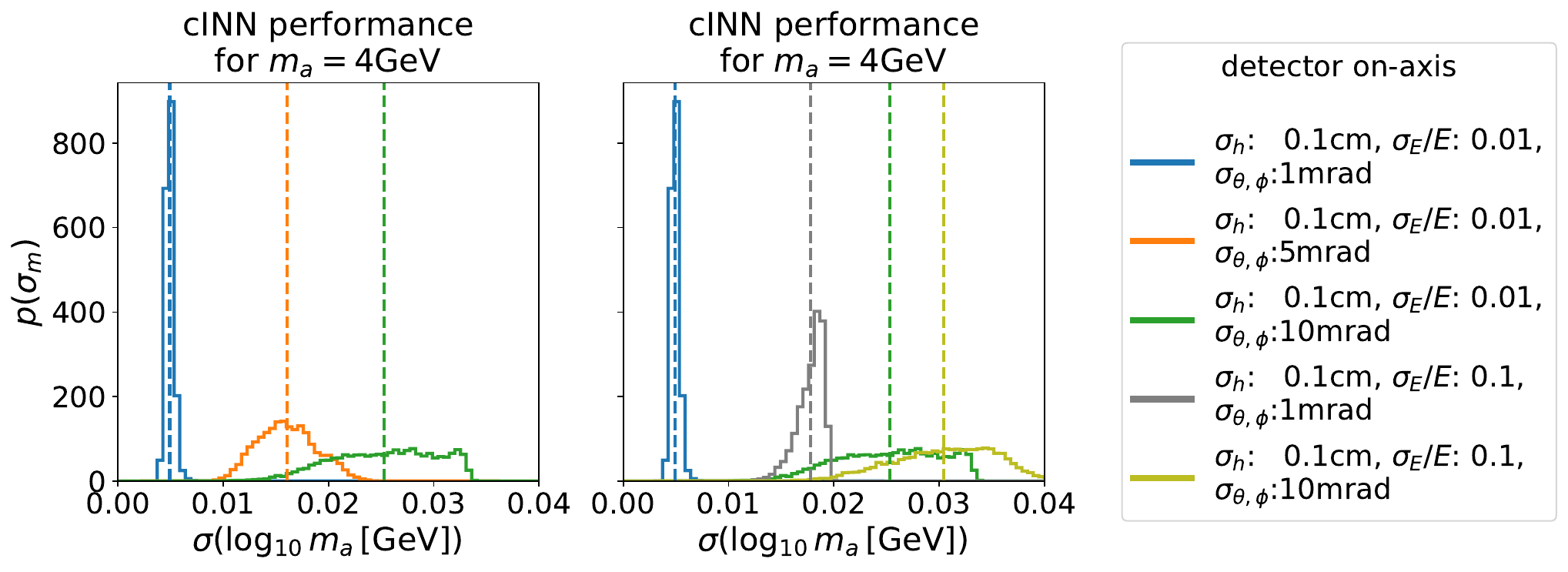}
    \caption{\label{fig:perf_angle_on_appendix}Same as \cref{fig:perf_angle_on,fig:perf_angen_on}, but for $m_a=\SI{0.2}{\giga\electronvolt}$ (top) and $m_a=\SI{4}{\giga\electronvolt}$ (bottom).}
\end{figure*}

\section{Further performance plots}\label{sec:appendix_plots}

\begin{figure*}[t]
    \centering
    \includegraphics[width=0.8\textwidth]{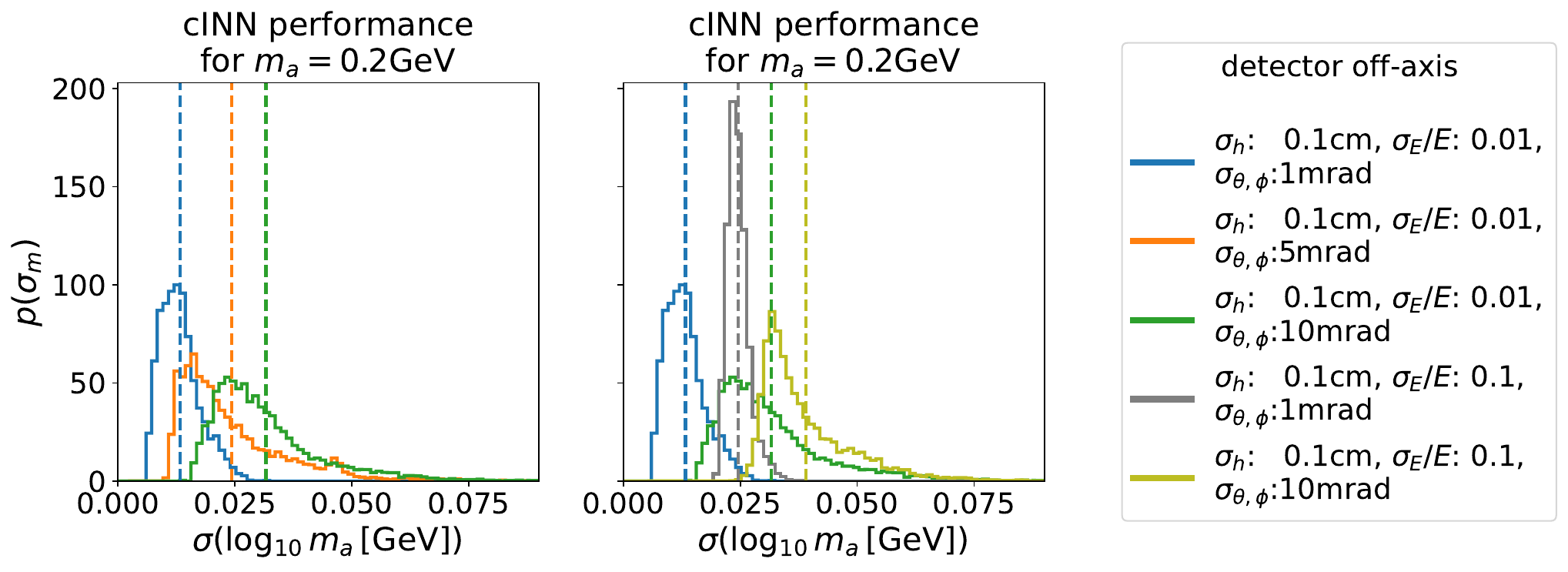}
    
    \vspace{2mm}
    \includegraphics[width=0.8\textwidth]{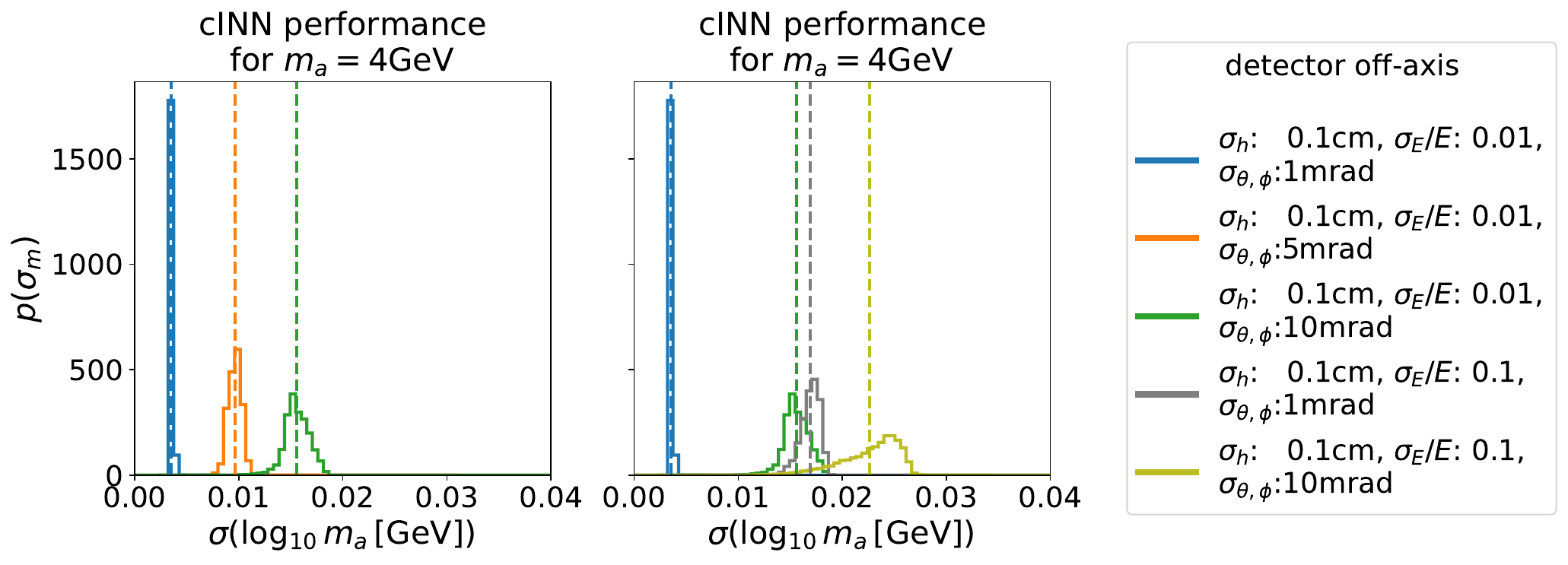}
    \caption{Same as \cref{fig:perf_angle_off}, but for $m_a=\SI{0.2}{\giga\electronvolt}$ (top) and  $m_a=\SI{4}{\giga\electronvolt}$ (bottom). \label{fig:perf_angle_off_appendix}}
\end{figure*}

In \cref{sec:performance_on,sec:performance_off} we have focused on the case of $m_a=\SI{1}{\giga\electronvolt}$. In \cref{fig:perf_angle_on_appendix,fig:perf_angle_off_appendix} we show the same distributions for the other benchmark masses of $\SI{0.2}{\giga\electronvolt}$ and $\SI{4}{\giga\electronvolt}$. While the same qualitative conclusions about the role of angular and energy resolution hold in these cases, their quantitative effects differ. It is also worth remembering that the cINN does not (and should not) combine the low-level observables in the same way for different detector setups. This implies that the shape of the $\sigma_m$ distributions can vary considerably when we change the detector resolutions.

\newpage

\section{Performance in the case of poor angular resolution}\label{sec:calo_hit}

\begin{figure*}
    \centering
    \includegraphics[width=\textwidth]{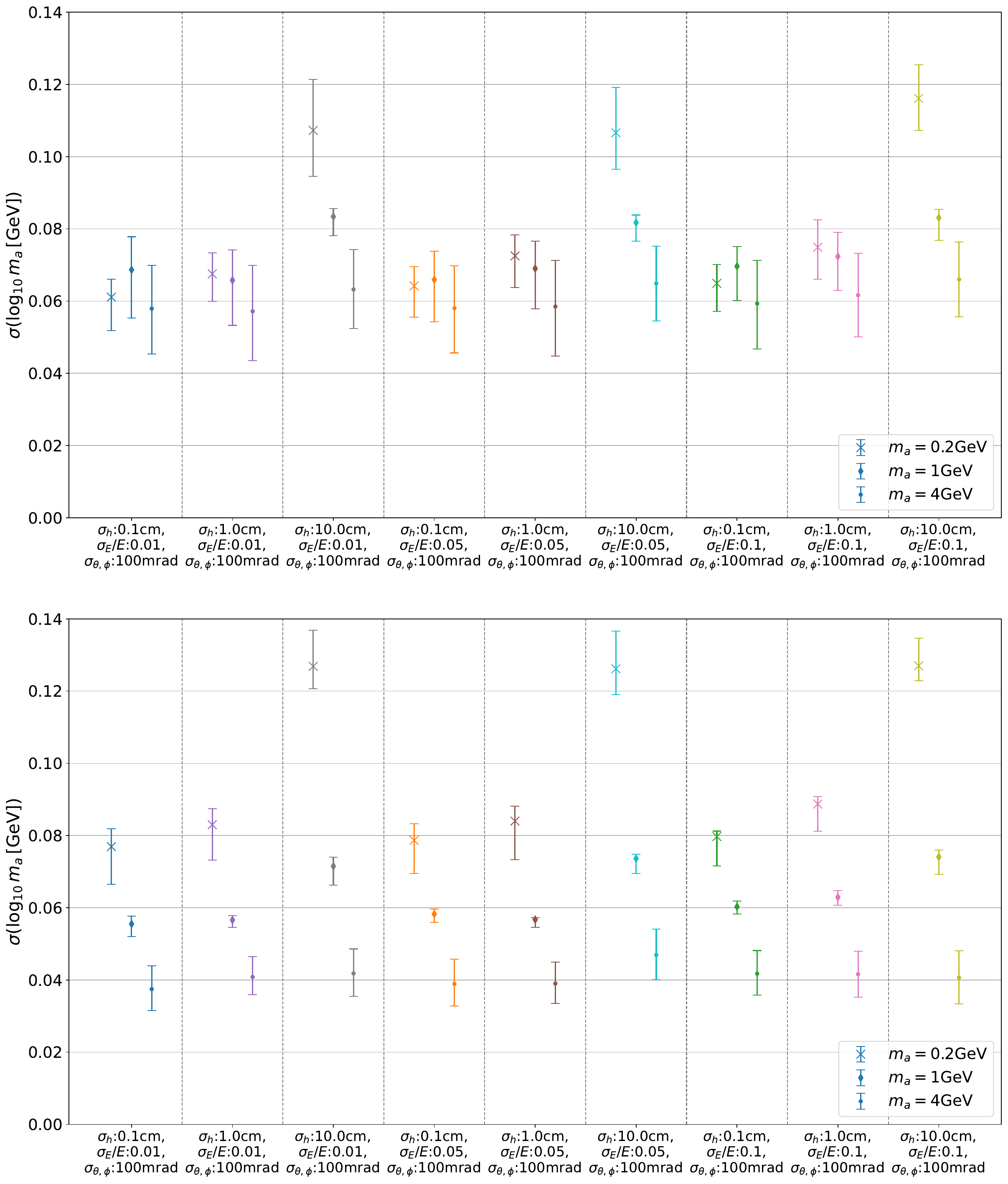}
    \caption{Compact performance comparison for fixed and large angular uncertainty for the case of on-axis (upper plot) and off-axis geometry (upper plot)). Different colors correspond to different detector setups and different markers are used for the benchmark masses. The vertical bands indicate the 25 and 75 percentile of the distribution.}
    \label{fig:compact_perf_comp_calo_displ}
\end{figure*}

In this appendix we show the results for the case $\sigma(\theta)=\sigma(\phi)=\SI{100}{\milli\radian}$ and varying calorimeter hit uncertainty $\sigma(h)$ (see \cref{tab:unc_setups_calohit}) for the on-axis and off-axis geometries in the upper and lower plots of \cref{fig:compact_perf_comp_calo_displ} respectively.

These performance plots are useful to illustrate two things we have mentioned in the main text. First, the calorimeter hit resolution becomes important only in the case of poor angular resolution as can be seen for both the on-axis and off-axis detectors. The effect is particularly relevant when the calorimeter hit resolution is low ($\sigma(h)\gtrsim \SI{10}{\centi\meter}$). Second, the energy resolution is more important when the angular resolution is good ($\sigma(\theta)=\sigma(\phi)\lesssim \SI{10}{\milli\radian}$). In the cases portrayed here of poor angular resolution, we see that having good energy resolution does not help in inferring the ALP mass.

\begin{table}[h]
    \centering
    \begin{tabular}{c|c}
        Feature uncertainty & Values scanned \\
        \hline
          $\sigma(E)/E$ & $[0.01, 0.05, 0.1]$ \\  
          $\sigma(h)$ & $[\SI{0.1}{\centi\meter}, \SI{1}{\centi\meter}, \SI{10}{\centi\meter}]$ \\
          $\sigma(\theta),\sigma(\phi)$ & $[\SI{100}{\milli\radian}]$ 

    \end{tabular}
    \caption{Summary of the detector setups considered in this appendix. Each uncertainty value is combined with each of the other uncertainty values for a total of 9 combinations. These combinations focus on the case of poor angular resolutions.}
    \label{tab:unc_setups_calohit}
\end{table}

\bibliographystyle{bibstyle}
\bibliography{bibliography}

\end{document}